\documentclass{article}

\usepackage[utf8]{inputenc}
\usepackage[T1]{fontenc}
\usepackage[english]{babel}
\usepackage{lmodern}
\usepackage{soul}
\usepackage{xcolor}

\usepackage{graphicx}
\usepackage{subcaption}
\usepackage{tikz}

\usepackage[nocompress]{cite} 
\usepackage{etoolbox}
\AtBeginEnvironment{thebibliography}{%
  \footnotesize
  \setlength{\itemsep}{0pt}%
  \setlength{\parsep}{0pt}%
  \setlength{\parskip}{0pt}%
  \setlength{\topsep}{0pt}%
}
\usepackage{hyperref}  
\hypersetup{
    colorlinks=true, 
    linkcolor=black, 
    filecolor=magenta,      
    urlcolor=blue,
    citecolor=blue,
}

\usepackage[obeyFinal]{todonotes}
\usepackage{comment}
\usepackage[title]{appendix}
\usepackage{amsmath}  
\usepackage{amsfonts,dsfont}
\usepackage{amssymb}
\usepackage{amsthm}
\usepackage{mathtools}
\usepackage{thmtools}
\usepackage{bm} 
\usepackage{enumitem}

\numberwithin{equation}{section} 

\newlength{\bredde}
\def\slash#1{\settowidth{\bredde}{$#1$}\ifmmode\,\raisebox{.15ex}{/}
\hspace*{-\bredde} #1\else$\,\raisebox{.15ex}{/}\hspace*{-\bredde} #1$\fi}
\renewcommand{\epsilon}{\varepsilon} 
\renewcommand{\subset}{\subseteq} 
\let\Re\relax\DeclareMathOperator{\Re}{Re} 
\let\Im\relax\DeclareMathOperator{\Im}{Im} 

\newcommand*{\euler}{\mathrm{e}} 
\newcommand*{\iunit}{\mathrm{i}} 
\newcommand*{\dif}{\mathop{}\!\mathrm{d}} 

\def\C{\mathbb{C}}

\def\E{\mathbb{E}}

\def\R{\mathbb{R}}
\newcommand{\bfR}{\mathbf{R}}
\newcommand*{\Rset}{\mathbb{R}} 
\newcommand*{\Cset}{\mathbb{C}} 

\def\syme{\mathfrak{e}}

\def\val{\boldsymbol{\alpha}}
\def\vnu{\boldsymbol{\nu}}

\usepackage{extpfeil}
\newextarrow{\xrightrightarrows}{{5}{8}{0}{0}}
{\bigRelbar\bigRelbar{\bigtwoarrowsleft\rightarrow\rightarrow}}

\DeclareMathOperator{\erf}{erf} 
\DeclareMathOperator{\erfc}{erfc} 
\DeclareMathOperator{\Trace}{Tr} 
\DeclareMathOperator{\Pf}{Pf} 
\newcommand*{\cconj}[1]{\overline{#1}} 

\declaretheorem[numberwithin=section]{proposition}
\declaretheorem[numberlike=proposition]{theorem}
\declaretheorem[numberlike=proposition]{lemma}

\declaretheorem[numberlike=proposition]{remark}

\makeatletter
\newcommand*{\doublerightarrow}[2]{\mathrel{
  \settowidth{\@tempdima}{$\scriptstyle#1$}
  \settowidth{\@tempdimb}{$\scriptstyle#2$}
  \ifdim\@tempdimb>\@tempdima \@tempdima=\@tempdimb\fi
  \mathop{\vcenter{
    \offinterlineskip\ialign{\hbox to\dimexpr\@tempdima+1em{##}\cr
    \rightarrowfill\cr\noalign{\kern.5ex}
    \rightarrowfill\cr}}}\limits^{\!#1}_{\!#2}}}
\makeatother

\AtEndDocument{%
  \vspace{2em}
  \noindent\rule{0.4\textwidth}{0.4pt}\par
  \vspace{0.6em}
  \footnotesize
  \noindent
  \begin{tabular}{@{}ll@{}}
    \textsc{Can Atacan}   & \href{mailto:catacan@math.uni-bielefeld.de}{\texttt{catacan@math.uni-bielefeld.de}} \\
    \textsc{Noah Ayg\"un} & \href{mailto:noah.ayguen@uni-bielefeld.de}{\texttt{noah.ayguen@uni-bielefeld.de}} \\
    \textsc{Yong-Woo Lee} & \href{mailto:hellowoo@snu.ac.kr}{\texttt{hellowoo@snu.ac.kr}}
  \end{tabular}
}

\begin{document}
\title{
Scaling limits for products of induced symplectic Ginibre matrices
}
\author{{\sc Can Atacan$^{1,2}$, Noah Ayg\"un$^{1,2}$, Yong-Woo Lee$^{1,2}$}\\~\\
$^1$Faculty of Physics, Bielefeld University, PO-Box 100131, D-33501 Bielefeld, Germany\\
$^2$Department of Mathematical Sciences,
Seoul National University, Seoul 151-747,\\ Republic of Korea\\
}


\maketitle

\begin{abstract}

We study the local eigenvalue correlations of products of independent induced symplectic Ginibre matrices in the large-dimensional limit. We establish the scaling limits of the correlation functions in the bulk, at the spectral edges, and outside the limiting spectral support, both near and away from the real axis. At regular spectral points, we show that the limiting correlations coincide with those of a single symplectic Ginibre ensemble, exhibiting Pfaffian structures near the real axis and determinantal structures away from it. At the origin, where the limiting spectral density develops a singularity, we derive an integral representation of the parameter-dependent limiting Pfaffian prekernel. Our proofs are based on a unified approach involving a differential equation for the skew prekernels and their asymptotics.
\end{abstract}


\section{Introduction and main results}\label{intro}
Products of random matrices have long been studied in probability theory and mathematical physics, dating back to the early development of random matrix theory. Classical investigations, notably the pioneering works of Bellman \cite{Bel54}, Furstenberg and Kesten \cite{FK60}, and Oseledec \cite{Ose68}, primarily concerned the asymptotic behavior of long products of matrices of fixed dimension. More recently, considerable attention has been directed toward the spectral properties of products of large-dimensional random matrices. Since matrix multiplication does not generally preserve Hermiticity, even when the individual factors are Hermitian, the spectral analysis of random matrix products naturally leads to non-Hermitian random matrix theory. We refer to the surveys \cite{AI2015, BF25} for further historical background and applications of random matrix products.

The Ginibre ensembles stand out as a foundational model for studying non-Hermitian random matrices. At the global scale, matrix multiplication can substantially deform the limiting eigenvalue distribution \cite{AB12, BJW10, OS11}; see also \cite{GT11}. A natural question is whether such changes in the behavior of the global spectra are also reflected in the local correlations between eigenvalues. For products of complex Ginibre matrices, which form determinantal point processes, it is known that the local correlations in the bulk (away from the origin) and at the edge remain universal, coinciding with those of a single complex Ginibre matrix, whereas distinct, parameter-dependent limits arise at the origin \cite{AB12, LW16}.
This contrast between regular spectral points and the origin raises the question of whether analogous phenomena persist in other symmetry classes.
In the real symmetry class, real eigenvalue statistics for products of Ginibre matrices have been studied in \cite{FI16, FS23}, whereas the symplectic counterpart remains less explored.

\begin{figure}[t]
    \begin{center}
    \begin{subfigure}{0.32\textwidth}
        \begin{center}
            \includegraphics[width=\linewidth]{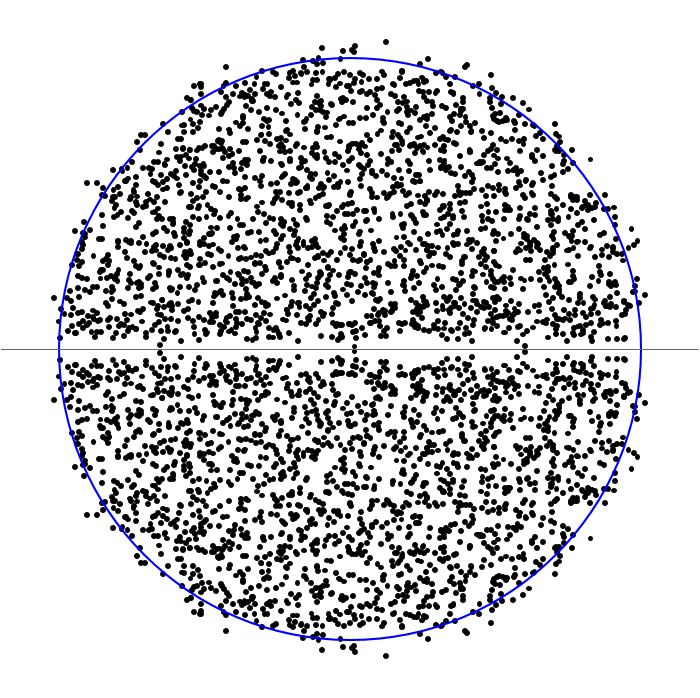}
        \end{center}
        \subcaption{$m=1$, $\nu_1 = 0$. }
    \end{subfigure}
    \begin{subfigure}{0.32\textwidth}
        \begin{center}
            \includegraphics[width=\linewidth]{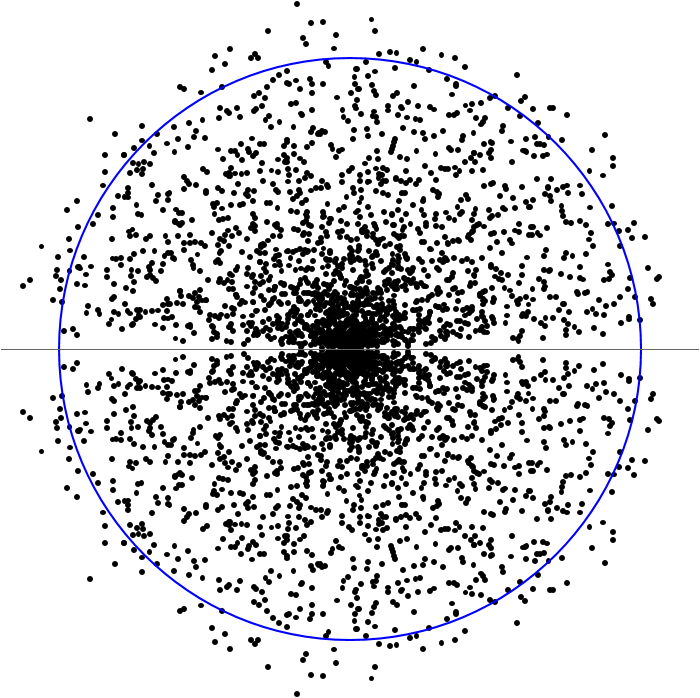}
        \end{center}
        \subcaption{$m=3$, $(\nu_1,\nu_2,\nu_3) = (1,3,5)$. } 
    \end{subfigure}
    \begin{subfigure}{0.32\textwidth}
        \begin{center}
            \includegraphics[width=\linewidth]{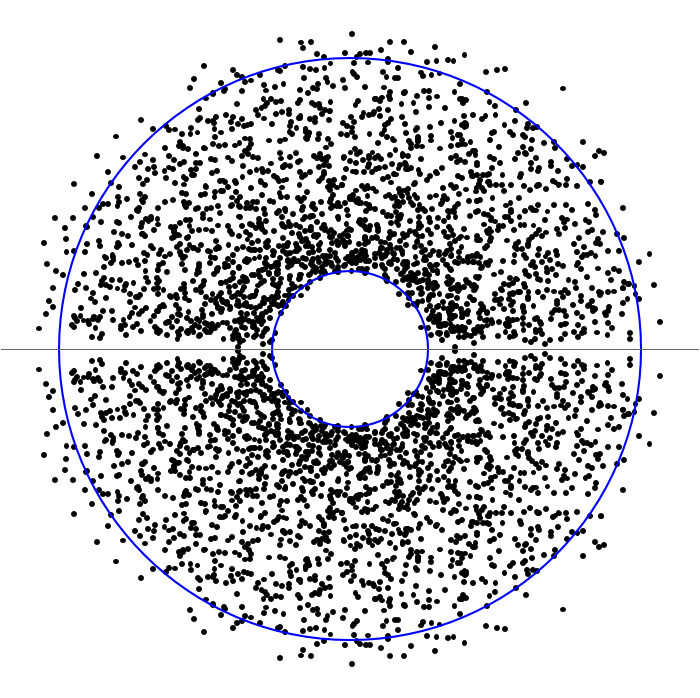}
        \end{center}
        \subcaption{$m=3$, $(\nu_1,\nu_2,\nu_3) = (50,50,80)$.  } 
    \end{subfigure}
    \captionsetup{width=0.96\textwidth}
    \caption{Spectra of $m$ products of induced symplectic Ginibre ensembles with $N=80$, obtained from $25$ samples. The solid blue circles mark the boundary of the rescaled limiting droplet $S$ given by \eqref{eq:macroscopic-radii}.}
    \label{fig:examples}
    \end{center}
\end{figure}

In the symplectic symmetry class, the eigenvalues occur in complex-conjugate pairs, giving rise to repulsion from the real axis, as illustrated in Figure~\ref{fig:examples}. Consequently, the local eigenvalue statistics exhibit distinct behavior near and away from the real axis. For a single symplectic Ginibre matrix, the real bulk scaling limit dates back to \cite{Mehtabook}, while the corresponding real edge limits were established in \cite{ABK22, AP14, Ly21}.
Integrable structures and real-axis scaling limits have been extensively studied across a broad family of exactly solvable symplectic random matrix models, which includes elliptic \cite{Kan02, BE23, BES23}, induced \cite{FBKSZ12, BC23}, spherical \cite{May13, MP17, BF23}, and truncated ensembles \cite{KL21}.
These developments have revealed common Pfaffian correlation structures near the real axis, providing a largely unified picture of local universality among the classical integrable symplectic ensembles. Away from the real axis, the correlations are expected to reduce to the determinantal structure of the complex Ginibre ensemble. However, despite this widely anticipated behavior and its verification in certain models and regimes, a general rigorous understanding of the convergence of correlation functions away from the real axis remains incomplete, even for a single symplectic Ginibre matrix. Thus, while considerable progress has been made in understanding universality along the real line, its complex-bulk counterpart remains less developed.

Against this background, products of symplectic Ginibre matrices stand out as a notable missing case in the local universality theory of exactly solvable symplectic ensembles. As for a single matrix, their eigenvalue correlations admit a Pfaffian representation in terms of skew-orthogonal polynomials, providing an algebraic framework for investigating their local correlation structure. Previous studies have derived explicit finite-dimensional correlation structures \cite{Ips13, IK14, AIS14, Ips15} and investigated their asymptotic behavior, particularly at the origin, where the limiting correlations depend on the number of matrix factors and the induced parameters \cite{Ips13, Ips15}. At regular spectral points, however, the local scaling limits have not yet been rigorously established.

In this paper, we investigate products of induced symplectic Ginibre matrices and establish the scaling limits of their local eigenvalue correlation functions across all local spectral regimes. We show that the local correlations in the bulk and at the edge converge to the universal Pfaffian kernels near the real axis and to the determinantal kernels of the complex Ginibre ensemble away from it. At the origin, we obtain distinct microscopic limits that depend explicitly on the number of matrix factors and the induced parameters. We also describe the limiting behavior outside the spectral support. Together, these results provide a unified description of the local eigenvalue statistics of symplectic Ginibre products, showing that, despite the deformation of the global eigenvalue distribution under multiplication, the local correlations at regular spectral points retain the universality observed for a single matrix. We now introduce the model and state our main results.

\bigskip

We consider the product of $m$ independent induced symplectic Ginibre matrices
\begin{equation}
    Y_N:=X_m\cdots X_1,
\end{equation}
where each $X_\ell$ is an $N\times N$ quaternion random matrix with probability density proportional to
\begin{align}
    \det (X_\ell^{\dagger} X_\ell)^{\nu_\ell}\euler^{-\Trace X_\ell^{\dagger }X_\ell}.
\end{align}
Here $\nu_\ell$ are the induced parameters, and the determinant and trace are understood in the standard $2N\times 2N$ complex representation of quaternion matrices.
When each $\nu_\ell$ is a nonnegative integer, the matrix $X_\ell$ can be constructed from an $(N+\nu_\ell)\times N$ rectangular quaternion Ginibre matrix via polar decomposition. Accordingly, products of induced square matrices are closely related to products of rectangular Ginibre matrices. Alternatively, at the level of eigenvalue distributions, the induced ensemble can be interpreted in terms of a larger square symplectic Ginibre ensemble conditioned to have eigenvalue pairs pinned at the origin. We refer to \cite{FBKSZ12, IK14, AIS14}, \cite[Section 2.4]{For16}, and \cite[Section 11.2]{BF25} for further details and related constructions.

The focus of this paper is the large-dimensional limit, keeping the number of factors $m$ fixed and allowing the induced parameters to grow linearly with the dimension.
More precisely, we assume throughout the paper that $\nu_\ell > -\frac{1}{2}$, and as $N\to\infty$ that
\begin{equation} \label{eq: nu parameter assumption}
    \nu_\ell=\alpha_\ell N+O(1),
\end{equation}
for fixed constants $\alpha_\ell\geq 0$, for all $\ell=1,\ldots, m$.

Our primary objects of interest are the $n$-point eigenvalue correlation functions of $Y_N$.
For $1 \leq n \leq N$ and $\zeta_1, \ldots, \zeta_n \in \C$ satisfying $\zeta_j \neq \zeta_k, \cconj{\zeta}_k$ for $j \neq k$, they are defined by
\begin{equation} \label{eq:correlation-definition}
    \bfR_{N,n}(\zeta_1,\ldots,\zeta_n)
    :=
    \lim_{\epsilon \to 0}\frac{1}{(\pi\epsilon^2)^n} \E_N \bigg[ \prod_{j=1}^n \mathcal{N}\big(D(\zeta_j, \epsilon)\big) \bigg],
\end{equation}
where $\mathcal{N}(A)$ is the number of eigenvalues of $Y_N$ inside a set $A \subset \C$, and $D(\zeta, r)$ denotes the disk of radius $r$ centered at $\zeta$.
Here, $\E_N$ denotes the expectation with respect to the law induced from that of $Y_N$.
Passing from the matrix distribution of $Y_N$ to an explicit Pfaffian representation for the $n$-point function combines a generalized Schur decomposition with skew-orthogonal polynomials and de Bruijn's integration formula; see \cite{Ips13, IK14} and \cite[Chapter~4]{Ips15} for the full derivation. More precisely, for any $1\leq n\leq N$, we have
\begin{equation}
\bfR_{N,n}(\zeta_1,\ldots,\zeta_n)
=
\prod_{j=1}^n(\bar\zeta_j-\zeta_j)\omega_{\vnu}(\zeta_j)
\Pf
\left[
\begin{array}{cc}
\varkappa_N(\zeta_j,\zeta_k) & \varkappa_N(\zeta_j,\bar\zeta_k)\\[2pt]
\varkappa_N(\bar\zeta_j,\zeta_k) & \varkappa_N(\bar\zeta_j,\bar\zeta_k)
\end{array}
\right]_{j,k=1}^n,
\label{eq:finite-pfaffian}
\end{equation}
where
\begin{equation} \label{eq. weight function omega m}
    \omega_{\vnu}(\zeta) :=
    G^{m,0}_{0,m} \left( 2^m|\zeta|^2 \,\middle|\,
    \begin{matrix}
        -
        \\
        2\nu_1,\ldots,2\nu_m
    \end{matrix}
    \right).
\end{equation}
Here, the function $G_{0,m}^{m,0}$ denotes the Meijer $G$-function given in \eqref{eq: def Meijer-G}, and
\begin{equation}\label{eq:finite-prekernel}
    \varkappa_N(\zeta,\eta) := 2^m\pi^{\frac m2-1} \sum_{k=0}^{N-1}\sum_{j=0}^{k} \frac{ \zeta^{2k+1}\eta^{2j}-\eta^{2k+1}\zeta^{2j} }{ \prod_{\ell=1}^m 4^{\nu_\ell} \Gamma(\nu_\ell+j+1)\Gamma(\nu_\ell+k+3/2) }.
\end{equation}

The one-point function encodes the mean eigenvalue density, which, in the large-dimensional limit, determines the spectral regions where the different local regimes arise.
The eigenvalue statistics of symplectic Ginibre ensembles admit a two-dimensional Coulomb gas interpretation \cite{KS99, For16}, in which the eigenvalues interact logarithmically with one another and with their complex-conjugate counterparts. At the global scale, this interpretation allows the limiting spectral distribution to be described through logarithmic potential theory \cite{BC12}; see also \cite[Section 10.2]{BF25}.
Under the scaling in \eqref{eq: nu parameter assumption}, the eigenvalues of $Y_N$ are typically of order $N^{m/2}$. After rescaling by this factor, their limiting macroscopic distribution is supported on the annulus: \cite[Eq.~(44)]{Ips13}, \cite[Theorem~1.2]{LW16}
\begin{equation} \label{eq:macroscopic-radii}
    S := \Big\{p \in \C: r_{\rm in} \leq |p| \leq r_{\rm out} \Big\},
    \qquad
    r_{\rm in} := \prod_{\ell=1}^{m}\alpha_\ell^{1/2},
    \qquad
    r_{\rm out} := \prod_{\ell=1}^{m}(1+\alpha_\ell)^{1/2}.
\end{equation}
To describe the limiting density, for any $p \in \C$, let $\lambda_{\val}(p)$ be the largest real root of
\begin{equation}\label{eq:lambda-root}
    \prod_{\ell=1}^{m}(\lambda_{\val}(p)+\alpha_\ell)-|p|^2=0.
\end{equation}
Then, the rescaled one-point function satisfies
\begin{equation*}
    \lim_{N\to \infty} N^{m-1} \bfR_{N,1}(N^{m/2} p) = \frac{2}{\pi} \rho(p) \, \mathbf{1}_S(p)
\end{equation*}
for any $p \in \C\setminus\R$ away from the spectral boundary $\partial S$, where
\begin{equation} \label{eq: def rho macroscopic density}
    \rho(p) :=  \frac{1}{ |p|^2} \Big( \sum_{l=1}^m \frac{1}{\lambda_{\val}(p)+\alpha_l} \Big)^{-1}.
\end{equation}
On the real axis, the finite-dimensional one-point function vanishes, reflecting the repulsion between conjugate eigenvalues.
In the special case $\alpha_1 = \cdots = \alpha_m = 0$, the limiting density profile simplifies to the power-law density found in \cite{AB12, BJW10},
\begin{equation*}
    \rho(p) \Big|_{\alpha_1 = \cdots = \alpha_m = 0}=\frac{1}{m} |p|^{-2 + 2/m}.
\end{equation*}
For $m>1$, this density develops a power-law singularity at the origin, distinguishing it from the regular bulk and motivating a separate analysis of the microscopic correlations near zero.

\begin{figure}[t]
    \centering
    \begin{tikzpicture}

        \node[anchor=south west, inner sep=0] (image)
            at (0,0) {
                \includegraphics[width=0.63\linewidth]
                {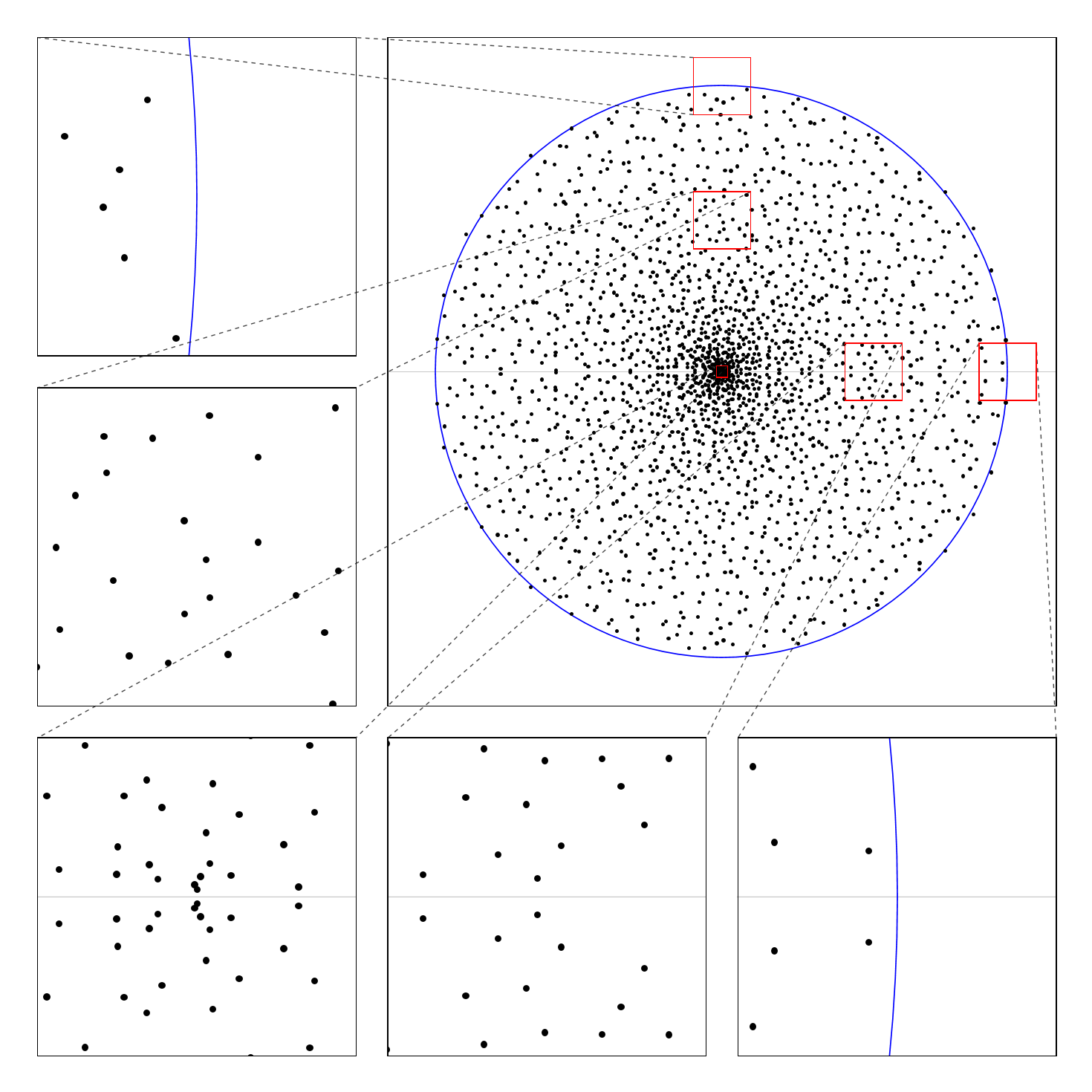}
            };

        \begin{scope}[
            x={(image.south east)},
            y={(image.north west)}
        ]

        \tikzset{
            regime/.style={
                anchor=north west,
                align=left,
                font=\scriptsize,
                fill=white,
                inner sep=2pt
            }
        }

        \node[regime] at (0.045,0.96) {
            \textbf{Complex edge}\\
            Theorem~\ref{thm:complex-line}~\ref{thm: complex edge}
        };

        \node[regime] at (0.045,0.64) {
            \textbf{Complex bulk}\\
            Theorem~\ref{thm:complex-line}~\ref{thm: complex bulk}
        };

        \node[regime] at (0.045,0.315) {
            \textbf{Origin}\\
            Theorem~\ref{thm:origin}
        };

        \node[regime] at (0.365,0.315) {
            \textbf{Real bulk}\\
            Theorem~\ref{thm:real-line}~\ref{thm: real bulk}
        };

        \node[regime] at (0.685,0.315) {
            \textbf{Real edge}\\
            Theorem~\ref{thm:real-line}~\ref{thm: real edge}
        };

        \end{scope}
    \end{tikzpicture}
    \captionsetup{width=0.7\linewidth}

    \caption{
        Nontrivial scaling regimes for local
        eigenvalue correlation functions.
        Here, a single realization of $Y_N$ with $N=1000$, $m=2$, $\nu_1=\nu_2=0$ is used.
    }
    \label{fig: scaling regimes}
\end{figure}

\medskip

The geometry of the limiting spectral support, together with the conjugation symmetry of the eigenvalues, naturally leads to five distinct scaling regimes for the local correlation functions. Away from the origin, four regimes arise according to whether the reference point lies on or away from the real axis, and whether it belongs to the interior or a regular boundary of the limiting support. These are referred to as the \emph{real bulk}, \emph{real edge}, \emph{complex bulk}, and \emph{complex edge} regimes. Along the real axis, the repulsion between complex-conjugate eigenvalues remains visible on microscopic scales, leading to the universal Pfaffian correlation structures characteristic of planar symplectic ensembles. Away from the real axis, by contrast, the separation of conjugate eigenvalues suggests that the local correlations should converge to those of the complex Ginibre determinantal process. Establishing this expected determinantal limit rigorously is one of the main results of the present work.

The fifth regime concerns the origin, which requires separate consideration due to the singular behavior that arises from matrix multiplication. Unlike the regular spectral regimes, the microscopic correlations near the origin retain an explicit dependence on both the number of matrix factors and the induced parameters. To capture these parameter-dependent limits, we consider the origin regime with $\nu_1,\ldots,\nu_m$ fixed, so that $\alpha_\ell=0$ in \eqref{eq: nu parameter assumption}. We also examine reference points outside the limiting spectral support, where the local correlation functions vanish in the large-dimensional limit. The five nontrivial scaling regimes, together with their corresponding main results, are illustrated in Figure~\ref{fig: scaling regimes}.

\medskip

We first consider the universal scaling limits in the regular bulk and at the spectral edges.
Let $p\neq0$ be a fixed reference point. To investigate the local eigenvalue statistics near $N^{m/2}p$, we introduce the local coordinates
\begin{equation} \label{eq. micro scale zeta eta z w}
    \zeta_j = N^{m/2}\Big(p + \frac{ \euler^{\iunit\theta}z_j}{\sqrt{N\rho(p) }}\Big),
\end{equation}
where $\theta\in[0,2\pi)$ denotes the outward normal angle if $p$ is on a regular boundary of $S$, and $\theta=0$ otherwise.
We then define the rescaled $n$-point correlation function from \eqref{eq:correlation-definition} by
\begin{equation} \label{eq: def rescaled correlation}
    R_{N,n}(z_1, \ldots, z_n) := \frac{1}{(N^{1-m}\rho(p))^n} \bfR_{N,n}(\zeta_1, \ldots, \zeta_n).
\end{equation}

We are now ready to state our first main result, establishing the bulk and edge scaling limits on the real axis.

\begin{theorem}[\textbf{Scaling limits near the real line}] \label{thm:real-line}
    Let $p \in \R \setminus \{0\}$ be fixed, and assume \eqref{eq: nu parameter assumption} holds.
    Then, for any fixed positive integer $n$, we have
    \begin{equation} \label{eq: Thm limiting n-point function real bulk and edge}
        \lim_{N \to \infty} R_{N,n}(z_1, \ldots, z_n) = \prod_{j=1}^n (\cconj{z}_j - z_j) \Pf\left[ \euler^{-|z_j|^2 -|z_k|^2} \begin{pmatrix}
            \kappa(z_j,z_k) & \kappa(z_j, \overline{z}_k) \\[2pt]
            \kappa(\overline{z}_j,z_k) & \kappa(\overline{z}_j, \overline{z}_k)
        \end{pmatrix}
        \right]_{j,k=1}^n
    \end{equation}
    locally uniformly for $z_1, \ldots, z_n \in \C$, where the limiting prekernel $\kappa(z,w)$ is given in each regime as follows.
    \begin{enumerate}[label=\textup{(\roman*)}]
        \item \textup{\textbf{The real bulk}:} \label{thm: real bulk}
        If $|p| \in (r_{\rm in}, r_{\rm out})$, then we have $\kappa(z,w) = \kappa_{\rm rb}(z,w)$, where
        \begin{equation} \label{eq. Thm 1 real bulk}
            \kappa_{\rm rb}(z,w) := \frac{2}{\sqrt{\pi}} \euler^{z^2+w^2} \erf(z-w).
        \end{equation}
        
        \item \textup{\textbf{The real edge}:}  \label{thm: real edge}
        If $|p|=r_{\rm out}$ or $|p|=r_{\rm in}>0$, then we have $\kappa(z,w) = \kappa_{\rm re}(z,w)$, where
        \begin{equation} \label{eq. Thm 1 real edge}
            \kappa_{\rm re}(z,w) := \frac{2}{\pi}\euler^{2zw} \int_{-\infty}^0 \euler^{-s^2} \sinh(2s(w-z)) \erfc((z+w)-s) \,\dif s.
        \end{equation}

        \item \textup{\textbf{The exterior}:} \label{thm: real exterior}
        If $|p| \notin [r_{\rm in}, r_{\rm out}]$, then we have $\kappa(z,w) = 0$.
    \end{enumerate}
\end{theorem}
A notable consequence of Theorem~\ref{thm:real-line} is that, after unfolding by the local density, the limiting correlations in the real bulk and at the real edge are independent of the number of matrix factors and the induced parameters.
In particular, they coincide with the universal limits of a single symplectic Ginibre matrix, despite the deformation of the macroscopic spectral distribution under multiplication.
The limiting prekernels also admit Wronskian representations, revealing an underlying integrable structure shared by several planar symplectic ensembles; see \cite{BES23, BF23}.

Our next result establishes the bulk and edge scaling limits in the complex plane away from the real axis.

\begin{theorem}[\textbf{Scaling limits away from the real line}] \label{thm:complex-line}
    Let $p \in \C \setminus \R$ be fixed, and assume \eqref{eq: nu parameter assumption} holds.
    Then, for any fixed positive integer $n$, we have
    \begin{equation}
        \lim_{N \to \infty} R_{N,n}(z_1, \ldots, z_n) = \det\begin{bmatrix}
            K(z_j,z_k)
        \end{bmatrix}_{j,k=1}^n
    \end{equation}
    locally uniformly for $z_1, \ldots, z_n \in \C$, where the limiting kernel $K(z,w)$ is given in each regime as follows.
    \begin{enumerate}[label=\textup{(\roman*)}]
        \item \textup{\textbf{The complex bulk}:}  \label{thm: complex bulk}
        If $|p| \in (r_{\rm in}, r_{\rm out})$, then we have $K(z,w) = K_{\rm cb}(z,w)$, where
        \begin{equation}
            K_{\rm cb}(z,w) := \frac{2}{\pi} \euler^{2z\cconj w-|z|^2 - |w|^2}.
        \end{equation}
        \item \textup{\textbf{The complex edge}:} \label{thm: complex edge}
        If $|p|=r_{\rm out}$ or $|p|=r_{\rm in}>0$, then we have $K(z,w) = K_{\rm ce}(z,w)$, where
        \begin{equation}
            K_{\rm ce}(z,w) := \frac{1}{\pi} \euler^{2z\cconj w-|z|^2 - |w|^2} \erfc(z+\cconj w). 
        \end{equation}

        \item \textup{\textbf{The exterior}:} \label{thm: complex exterior}
        If $|p| \notin [r_{\rm in}, r_{\rm out}]$, then we have $K(z,w) = 0$.
    \end{enumerate}
\end{theorem}
Theorem~\ref{thm:complex-line} shows that the local eigenvalue correlations away from the real axis converge to those of the complex Ginibre ensemble, independently of the number of matrix factors and the induced parameters. Thus, although the finite-dimensional correlations are Pfaffian, their scaling limits become determinantal at regular spectral points away from the real axis. As argued in \cite{AKMP19}, heuristics suggest that such a convergence arises by the separation of complex-conjugate eigenvalues away from the real axis, removing the self-conjugacy constraint on the local scale. For a single matrix, formal derivations in the complex bulk were given in \cite[Appendix~B]{AKMP19} and \cite{Ly21}. Related rigorous results have been obtained in other models and regimes, including the double-scaling edge limit \cite{ABK22} and the almost-circular regime of the induced symplectic Ginibre ensemble \cite{BC23}. Our theorem provides a unified, rigorous proof of this determinantal convergence in both the complex bulk and at the complex edges for products of induced symplectic Ginibre matrices at the level of all $n$-point correlation functions.

More broadly, Theorem~\ref{thm:complex-line} illustrates that away from the real axis, local eigenvalue correlations in the bulk and at regular spectral edges become independent of the underlying symmetry class.
Within the unitary class itself, the determinantal correlation structure governs single and product complex Ginibre matrices \cite{AB12, LW16} as well as general random normal matrix models \cite{AHM11, HW21, CW25}. Analogous off-axis reductions from Pfaffian to determinantal correlations have been established for the real Ginibre ensemble \cite{BS09, FN07, SW08}. These phenomena extend beyond integrable models to i.i.d. random matrices \cite{TV15, CES21, MO25, Os25}.

\medskip

Finally, we consider the scaling limits at the origin. Throughout this regime, we assume that the parameters $\nu_1,\ldots,\nu_m$ are fixed as $N\to\infty$. Unlike the regular bulk and edge regimes, the microscopic scaling at the origin does not require any $N$-dependent change of variables. As observed in \cite{Ips15}, the finite-dimensional prekernel \eqref{eq:finite-prekernel} converges locally uniformly to an infinite series as $N\to\infty$.

Our next result provides an integral representation of the origin scaling limit.

\begin{theorem}[\textbf{Scaling limits at the origin}]\label{thm:origin}
    Let $\vnu:=(\nu_1,\ldots, \nu_m)$ be fixed with $\nu_\ell > -\frac{1}{2}$ for all $\ell =1, \ldots, m$.
    Then, for any fixed positive integer $n$, we have
    \begin{equation}
    \label{def. origin n-point corelation function}
        R_n^{\vnu}(z_1, \ldots, z_n):=\lim_{N \to \infty} \bfR_{N,n}(z_1, \ldots, z_n) = \prod_{j=1}^n (\overline{z}_j - z_j) \omega_{\vnu}(z_j)  \Pf\begin{bmatrix}
            \kappa_{\rm ori}^{\vnu}(z_j,z_k) & \kappa_{\rm ori}^{\vnu}(z_j, \overline{z}_k) \\[2pt]
            \kappa_{\rm ori}^{\vnu}(\overline{z}_j,z_k) & \kappa_{\rm ori}^{\vnu}(\overline{z}_j, \overline{z}_k)
        \end{bmatrix}_{j,k=1}^n
    \end{equation}
    locally uniformly for $z_1, \ldots, z_n \in \C$, where
    \begin{align}\label{eq:origin-kernel}
    \begin{split}
        \kappa_{\rm ori}^{\vnu}(z, w) := \int_{[0,1]^m}
        &\left[ z\,{}_0F_{m-1} \Big(;1,\dots,1 \Big|z^2 \prod_{\ell=1}^m (1-t_\ell^2) \Big) -w\, {}_0F_{m-1} \Big(;1,\dots,1 \Big|w^2 \prod_{\ell=1}^m (1-t_\ell^2) \Big) \right] 
        \\
        & \quad \times I_{\rm ori}\Big(zw \prod_{\ell=1}^m t_\ell\Big) \prod_{\ell=1}^{m}t_\ell^{2\nu_\ell}\, \dif t_\ell,
    \end{split}
    \end{align}
    with ${}_0 F_{m-1}$ denoting the generalized hypergeometric function given in \eqref{eq: def p-F-q generalized hypergeometric}, and
    \begin{equation}\label{eq: def I-ori}
     I_{\rm ori}(z)
     :=\frac{4^m}{\pi}
     \sum_{k=0}^{\infty}
     \frac{(2^m z)^{2k}}
     {\prod_{j=1}^m\Gamma(2\nu_j+2k+1)}.
    \end{equation}
\end{theorem}
Theorem~\ref{thm:origin}  extends several known results for single and two-matrix symplectic ensembles.
When $m=1$, the hypergeometric function ${}_0F_0$ reduces to the exponential function, while $I_{\rm ori}$ can be expressed in terms of the two-parameter Mittag--Leffler function. The resulting representation agrees with the origin kernels studied in \cite{ABK22, BF23}.
In particular, for $\nu_1=0$, it recovers the real bulk scaling limit $\kappa_{\rm rb}(z,w)$ in \eqref{eq. Thm 1 real bulk}.
 Another notable special case is $m=2$ with $(\nu_1,\nu_2)=(0,\nu)$. Here ${}_0F_1$ reduces to a modified Bessel function, and the limiting prekernel can be identified with that of the chiral symplectic Ginibre ensemble; see \cite{Ak05} and \cite[Remark 4.9]{AEP22}. Thus, the representation \eqref{eq:origin-kernel} places these previously studied kernels within a common framework for arbitrary products.

The origin scaling limits for products with different numbers of factors are also related through a reduction formula. 
More precisely, for $m \geq 2$ with $\hat\vnu:=(\nu_{1},\ldots, \nu_{m-1})$ fixed, we have
\begin{equation}
    \lim_{\nu_m \to \infty }\nu_m^n R_n^{\vnu}(\sqrt{\nu_m }z_1, \ldots, \sqrt{\nu_m} z_n) = R_n^{\hat\vnu}(z_1, \ldots, z_n).
\end{equation}
In other words, after appropriate rescaling, the origin correlations of an $m$-fold product reduce to those of an $(m-1)$-fold product as one of the induced parameters tends to infinity. We refer to Proposition~\ref{prop. origin reduction} for the precise statement and proof.
A similar reduction formula is known for products of complex Ginibre matrices; see \cite[Proposition~4.11]{Ips15}.

\smallskip

We next briefly discuss the proof strategy, leaving further details to Section~\ref{sec: outline of the proof}.
As mentioned before, we employ a unified approach to establish the main results.
The main idea is to derive and analyze a differential equation satisfied by the prekernel.
Due to its skew-symmetry, substantial cancellations arise in the double sum in \eqref{eq:finite-prekernel} under the unfolding \eqref{eq. micro scale zeta eta z w}, which are difficult to handle directly. After applying a suitably chosen differential operator, however, the difficulty transfers to analysis of inhomogeneous terms, which are closely related to the correlation kernels of products of complex Ginibre matrices. This connection allows us to exploit their asymptotic behavior in the analysis of the Pfaffian correlations. The choice of the differential operator is, however, far from automatic and constitutes an essential step in the argument. An unsuitable choice may lead to a differential equation whose asymptotic analysis is no simpler than that of the original prekernel; see \cite[Section 6.2]{MP17} and \cite[Remark 1.2]{BF23} for further discussion.

An additional difficulty stems from the nontrivial macroscopic deformation induced by matrix multiplication. Unlike the single Ginibre ensemble, the local density $\rho(p)$ is expressed implicitly through the root $\lambda_{\val}(p)$ of the $m$-th order algebraic equation \eqref{eq:lambda-root}. Consequently, the asymptotic analysis involves quantities that depend intricately on the number of matrix factors, the induced parameters, and the reference point. A crucial step is to show that, after unfolding by the local density \eqref{eq. micro scale zeta eta z w}, these model-dependent contributions simplify to yield the universal limiting kernels. This requires a careful analysis of the rescaled differential equations in the different spectral regimes, particularly near the spectral boundaries.

The use of differential equations in the analysis of prekernels dates back to \cite{Mehtabook, Kan02, Ak05}, where the limiting correlations near the origin were investigated.
This approach was subsequently developed to establish the real bulk and edge scaling limits for various planar symplectic ensembles \cite{ABK22, AEP22, BE23, BES23, BF25, BN25}.
In the present work, we extend this method to the complex bulk and edge, as well as the exterior regimes, thereby establishing the scaling limits at regular spectral points within a unified framework. At the origin, the same approach leads to a higher-order differential equation, whose solution yields the integral representation in Theorem~\ref{thm:origin}.

\smallskip

Let us conclude this section by briefly mentioning several related directions.
Beyond local eigenvalue correlations, symplectic ensembles have been investigated from a variety of perspectives, including extreme eigenvalues and large deviations \cite{Ri03, Ri04, BLO26}, gap probabilities and radial symmetries \cite{APS09, Du21}, partition functions \cite{AB07, BKS23, Noda26a, Noda26b}; cf.~\cite[Proposition 10.8]{BF25}, counting statistics and spectral moments \cite{ABES23, ABO26}, and eigenvector overlaps \cite{AFK20, ABN26}. In the context of random matrix products, phase transitions when the number of factors grows with the dimension have also been studied in the complex Ginibre setting \cite{LW24}.
In the real symmetry class, Pfaffian structures and real eigenvalue statistics have likewise been investigated beyond the real Ginibre ensemble, including the real elliptic \cite{FN08} and spherical \cite{FM12} ensembles, products of real Ginibre matrices \cite{FI16, FS23}, and products of truncated orthogonal matrices \cite{FIK20, LMS22}.
This list is far from complete, and we refer to the comprehensive review \cite{BF25} for recent developments in this subject.

\subsection*{Organization of the paper}
The rest of the paper is organized as follows. In Section~\ref{sec:integrable}, we derive a finite-$N$ differential equation for the prekernel and outline the proof strategy for the main results. In Section~\ref{sec:asymptotic}, we establish the asymptotic estimates needed to analyze this differential equation. Combining these results, we prove Theorems~\ref{thm:real-line} and \ref{thm:complex-line} in Section~\ref{sec:proofs}.
Finally, Section~\ref{sec: Origin} is devoted to the scaling limits at the origin, including the proof of Theorem~\ref{thm:origin}.

\subsection*{Acknowledgments}
The authors are grateful to the DFG-NRF International Research Training Group IRTG
2235 supporting the Bielefeld-Seoul graduate exchange programme (NRF-2016K2A9A2A13003815). This work was partly funded by the German Research Foundation DFG with grant SFB 1283/2 2021 – 317210226 (CA, NA). 
The authors greatly appreciate Gernot Akemann and Sung-Soo Byun for suggesting this problem and for their helpful comments and support. NA is grateful for the hospitality of the Department of Mathematics at King’s College London.

\subsection*{AI use disclosure}
We used ChatGPT-6 for language editing and reviewing the proofs during the final stages of manuscript preparation. All mathematical arguments and proofs were independently developed and written by the authors, who take full responsibility for the content of this paper.

\section{Differential equation and outline of proof}\label{sec:integrable}
The key ingredient in our analysis is a differential equation satisfied by the finite-$N$ prekernel.
In this section, we derive this equation and explain how it is used to establish the local scaling limits.
We derive the differential equation in Section~\ref{sec: differential equation}, and then outline the proof strategy for the main results in Section~\ref{sec: outline of the proof}.

We begin by introducing some notation for the local rescaling.
Recalling the unfolding from \eqref{eq. micro scale zeta eta z w}, we write
\begin{equation} \label{def-xi^rho,p_N(z)}
    \xi_N^p(z) := N^{m/2} \Big(p + \frac{\euler^{\iunit \theta}z}{\sqrt{N\rho(p)}}\Big),
\end{equation}
for $z\in \Cset$, where $p$, $\rho(p)$, and $\theta$ are as in \eqref{eq. micro scale zeta eta z w}.
Then, for $p \in \C \setminus \{ 0 \}$, the rescaled $n$-point function \eqref{eq: def rescaled correlation} can be expressed as
\begin{equation} \label{eq. rescaled n-point function C}
    R_{N,n}(z_1, \ldots, z_n) = \prod_{j=1}^n \Big((\overline{p}-p)\sqrt{N\rho(p)} + \euler^{-\iunit \theta}\cconj{z}_j- \euler^{\iunit \theta} z_j\Big) \omega_{\vnu}(\xi_N^p(z_j))  \Pf\begin{bmatrix}
        \kappa_N^{1,1}(z_j,z_k) & \kappa_N^{1,2}(z_j, \overline{z}_k) \\[2pt]
        \kappa_N^{2,1}(\overline{z}_j,z_k) & \kappa_N^{2,2}(\overline{z}_j, \overline{z}_k)
    \end{bmatrix}_{j,k=1}^n,
\end{equation}
where the rescaled prekernels are defined by
\begin{align}
\begin{split} \label{eq. kappa-a,b (scaled prekernel)}
    \kappa_N^{1,1}(z,w) := \frac{\varkappa_N(\xi_N^p(z), \xi_N^p(w))}{(N^{1-m}\rho(p) )^{3/2}}, \qquad
    & \kappa_N^{1,2}(z,w) := \frac{\varkappa_N(\xi_N^p(z), \xi_N^{\cconj{p}}(w))}{(N^{1-m}\rho(p) )^{3/2}},
    \\
    \kappa_N^{2,1}(z,w) := \frac{\varkappa_N(\xi_N^{\cconj{p}}(z), \xi_N^p(w))}{(N^{1-m}\rho(p) )^{3/2}}, \qquad
    & \kappa_N^{2,2}(z,w) := \frac{\varkappa_N(\xi_N^{\cconj{p}}(z), \xi_N^{\cconj{p}}(w))}{(N^{1-m}\rho(p) )^{3/2}},
\end{split}
\end{align}
where $\varkappa_N$ is defined in \eqref{eq:finite-prekernel}.
In particular, when $p\in\mathbb R\setminus\{0\}$, the expression further simplifies to
\begin{equation} \label{eq: n-point function real line}
    R_{N,n}(z_1,\ldots,z_n)
=
\prod_{j=1}^n
\bigl(\cconj{z}_j-z_j\bigr)
\omega_{\vnu}\bigl(\xi_N^p(z_j)\bigr)
\Pf
\begin{bmatrix}
\kappa_N(z_j,z_k) & \kappa_N(z_j,\cconj{z}_k)\\[2pt]
\kappa_N(\cconj{z}_j,z_k) & \kappa_N(\cconj{z}_j,\cconj{z}_k)
\end{bmatrix}_{j,k=1}^n,
\end{equation}
where
\begin{equation} \label{eq. kappa (scaled prekernel)}
   \kappa_N(z,w) :=
    \euler^{-\iunit \theta}\frac{
    \varkappa_N\bigl(\xi_N^p(z),\xi_N^p(w)\bigr)
    }{
    (N^{1-m}\rho(p))^{3/2}
    }.
\end{equation}

We recall two special functions from \cite[Section 16]{NIST}: the generalized hypergeometric function and the Meijer $G$-function.
To define them, let $p$ and $q$ be nonnegative integers, and $a_1, \ldots, a_p$ and $b_1, \ldots, b_q$ be complex parameters.
The generalized hypergeometric function is then given by
\begin{equation} \label{eq: def p-F-q generalized hypergeometric}
    {}_p F_q(a_1, \ldots, a_p; b_1, \ldots, b_q| z) := \frac{\prod_{\ell=1}^q \Gamma(b_\ell)}{\prod_{j=1}^p \Gamma(a_j)} \sum_{k=0}^\infty \frac{\prod_{j=1}^p \Gamma(k+a_j)}{\prod_{\ell=1}^q \Gamma(k+b_\ell)} \frac{z^k}{k!},
\end{equation}
assuming none of $b_\ell$ is a non-positive integer.
Throughout the paper, we only consider the case $p \leq q$, for which the series is well-defined for any $z \in \C$.
Assume also that $m$ and $n$ are integers such that $0 \leq m \leq q$ and $0 \leq n \leq p$, and none of $a_k - b_j$ is a positive integer when $1 \leq k \leq n$ and $1 \leq j \leq m$.
The other is the Meijer $G$-function, which is defined through the Mellin--Barnes integral
\begin{equation} \label{eq: def Meijer-G}
    {G^{m,n}_{p,q}}\left(z \, \middle| \genfrac{}{}{0pt}{}{a_{1},\dots,a_{p}}{b_{1},\dots,b_{q}}\right)
    := \frac{1}{2\pi \iunit}\int_{L}{\frac{\prod_{\ell=1}^{m}\Gamma\left(b_{\ell}-s\right) \prod_{\ell=1}^{n}\Gamma\left(1-a_{\ell}+s\right)}{ \prod_{\ell=m}^{q-1}\Gamma\left(1-b_{\ell+1}+s\right)\prod_{\ell=n}^{p-1} \Gamma\left(a_{\ell+1}-s\right)}}z^{s}\,\dif s,
\end{equation}
where the integration path $L$ is chosen to separate the poles of the factors $\Gamma(b_\ell -s)$ from those of the factors $\Gamma(1-a_\ell+s)$, running from $-\iunit\infty$ to $\iunit\infty$ or looping around $\pm\infty$ so that the integral converges \cite[Section~16.17]{NIST}.
The generalized hypergeometric function is a special case of the Meijer $G$-function via 
\begin{equation} \label{eq: p-F-q and Meijer-G relation}
    {}_p F_q(a_1, \ldots, a_p; b_1, \ldots, b_q| z) = \frac{\prod_{\ell=1}^q \Gamma(b_\ell)}{\prod_{j=1}^p \Gamma(a_j)} {G^{1,p}_{p,q+1}}\left(-z \, \middle| \genfrac{}{}{0pt}{}{1-a_{1},\dots,1-a_{p}}{0, 1-b_{1},\dots,1-b_{q}}\right).
\end{equation}
We refer to \cite[Section 16]{NIST} and the references therein for further details.

\subsection{Differential equation} \label{sec: differential equation}

We derive the differential equation which will be the starting point of the subsequent analysis.
\begin{proposition}[\textbf{Differential equation for the prekernel}] \label{Prop. diff eq kappa}
    For any positive integer $N$ and $\zeta, \eta \in \C$ with $\zeta \neq 0$, the prekernel $\varkappa_N(\zeta,\eta)$ from \eqref{eq:finite-prekernel} satisfies the $m$-th order differential equation
    \begin{equation} \label{eq. diff eq finiteN-general case} 
        \Bigg[\frac{1}{\zeta} \prod_{l=1}^m\Big(\zeta\partial_\zeta +2\nu_l\Big) - 2^m \zeta\Bigg] \varkappa_N(\zeta,\eta) = I_{N,1}(\zeta,\eta) + I_{N,2}(\zeta,\eta) + I_{N,3}(\zeta,\eta),
    \end{equation}
    where the inhomogeneous terms are given by
    \begin{align} \label{def. inhom part I_1}
        I_{N,1}(\zeta,\eta) &:= \frac{2^{2m}}{\pi} \sum_{k=0}^{2N-1} \frac{(2^m \zeta\eta)^{k}}{\prod_{l=1}^m \Gamma(2\nu_l+k+1)}, \\
        \label{def. inhom part I_2}
        I_{N,2}(\zeta,\eta) &:= - \frac{ 2^{2m} \pi^{\frac{m}{2}-1} \zeta^{2N} }{\prod_{l=1}^m 4^{\nu_l} \Gamma(\nu_l + N + 1/2) } \sum_{j=0}^{N-1} \frac{\eta^{2j}}{ \prod_{l=1}^m \Gamma(\nu_l+j+1)}, \\ 
        \label{def. inhom part I_3}
        I_{N,3}(\zeta,\eta)  &:= - \frac{2^{2m} \pi^{\frac{m}{2}-1}}{\prod_{l=1}^m 4^{\nu_l} \Gamma(\nu_l)} \frac{1}{\zeta} \sum_{k=0}^{N-1} \frac{\eta^{2k+1}}{\prod_{l=1}^m \Gamma(\nu_l+k+3/2)}.
    \end{align}
    Here, we used the convention $1/\Gamma(0) = 0$.
\end{proposition}
\begin{proof}
    We decompose $\varkappa_N(\zeta,\eta) = F_N(\zeta,\eta) - F_N(\eta,\zeta)$, where
    \begin{equation} \label{eq. def of FN}
        F_N(\zeta,\eta) := c_m \sum_{k=0}^{N-1} \sum_{j=0}^k \frac{\zeta^{2k+1}\eta^{2j}}{\prod_{l=1}^m \Gamma(\nu_l +j+1) \Gamma(\nu_l+k+3/2)}, \qquad c_m := \frac{2^{m} \pi^{\frac{m}{2}-1}}{\prod_{l=1}^m 4^{\nu_l}}.
    \end{equation}
    Applying the differential operator $(\zeta \partial_\zeta + 2 \nu_l)$ repeatedly for $l= 1, \ldots, m$, and rearranging the summations, we obtain
    \begin{align}
        \Bigg[\prod_{l=1}^m\Big(\zeta\partial_\zeta +2\nu_l\Big)\Bigg] F_N(\zeta,\eta) &= 2^m c_m \sum_{k=0}^{N-1} \sum_{j=0}^k \frac{ \zeta^{2k+1}\eta^{2j}}{\prod_{l=1}^m \Gamma(\nu_l +j+1) \Gamma(\nu_l+k+1/2)}
        \nonumber \\
        &= 2^m \zeta^2 F_N(\zeta,\eta) + \zeta (I_{N,{\rm even}}(\zeta,\eta) + I_{N,2}(\zeta,\eta)),
        \label{eq. ODE F_N}
    \end{align}
    where 
    \begin{align*}
        I_{N,{\rm even}}(\zeta,\eta) &:= 2^m c_m \sum_{k=0}^{N-1} \frac{ \zeta^{2k}\eta^{2k}}{\prod_{l=1}^m \Gamma(\nu_l +k+1) \Gamma(\nu_l+k+\tfrac{1}{2})}.
    \end{align*}
    Similarly, for $F_N(\eta,\zeta)$, we have
    \begin{equation*}
        \Bigg[\prod_{l=1}^m\Big(\zeta\partial_\zeta +2\nu_l\Big)-2^m \zeta^2\Bigg] F_N(\eta,\zeta) = - \zeta (I_{N,{\rm odd}}(\zeta,\eta) + I_{N,3}(\zeta,\eta)),
    \end{equation*}
    where
    \begin{align*}
        I_{N,{\rm odd}}(\zeta,\eta) &:= 2^m c_m \sum_{k=0}^{N-1} \frac{ \eta^{2k+1}\zeta^{2k+1}}{\prod_{l=1}^m \Gamma(\nu_l +k+1) \Gamma(\nu_l+k+\tfrac{3}{2})}.
    \end{align*}
    Using the duplication formula for the Gamma function \cite[Eq. (5.5.5)]{NIST},
    we obtain
    \begin{align*}
        I_{N,{\rm even}}(\zeta,\eta) + I_{N,{\rm odd}}(\zeta,\eta) &= 2^m c_m \frac{\prod_{l=1}^m 4^{\nu_l}}{\pi^{\frac{m}{2}}}  \sum_{k=0}^{2N-1} \frac{(2^m \zeta\eta)^{k}}{\prod_{l=1}^m \Gamma(2\nu_l+k+1)} = I_{N,1}(\zeta,\eta).
    \end{align*}
    Combining the above and then dividing by $\zeta$, the desired differential equation follows.
\end{proof}

The inhomogeneous terms in \eqref{eq. diff eq finiteN-general case} converge to Meijer $G$-functions in the large-$N$ limit. More precisely, for fixed $\zeta, \eta \in \C$ with $\zeta \neq 0$ and fixed $\vnu$, it follows from \eqref{eq: def p-F-q generalized hypergeometric} and \eqref{eq: p-F-q and Meijer-G relation} that
\begin{align*}
    \lim_{N\to\infty} I_{N,1}(\zeta, \eta) &= \frac{2^{2m}}{\pi} G_{1,m+1}^{1,1}\left(-2^m \zeta \eta \middle|  \genfrac{}{}{0pt}{}{0}{0,-2\nu_1,\dots,-2\nu_m} \right),
    \qquad
    \lim_{N\to\infty} I_{N,2}(\zeta, \eta) =0,
    \\
    \lim_{N\to\infty} I_{N,3}(\zeta, \eta) &= - \frac{2^{2m} \pi^{\frac{m}{2}-1}}{\prod_{l=1}^m 4^{\nu_l} \Gamma(\nu_l)} \frac{\eta}{\zeta} G_{1,m+1}^{1,1}\left(-\eta^2 \middle|  \genfrac{}{}{0pt}{}{0}{0,-\nu_1-\frac{1}{2},\dots,-\nu_m-\frac{1}{2}} \right).
\end{align*}
Assuming $\zeta$ and $\eta$ to be bounded with at least one of $\nu_\ell$ being $0$, these give rise to the limiting differential equation considered in \cite[Eq.~(4.99)]{Ips15}.

We stress that the inhomogeneous part involves the determinantal correlation kernel of the corresponding unitary counterpart. In particular, up to normalization and rescaling, $I_{N,1}$ corresponds to the correlation kernel for products of the induced complex Ginibre ensemble; see, for instance, \cite{ARRS16, AB12, IK14}. Similar relations between Pfaffian and determinantal kernels arise in the study of other planar and spherical symplectic ensembles and their complex counterparts \cite{ABK22, BC23, BE23, BES23, BF23, BN25}.

For later use, we record that the differential operator in \eqref{eq. diff eq finiteN-general case} can be expanded in terms of symmetric polynomials.
More precisely, let $\syme_k(x_1, \ldots, x_n)$ be the $k$-th elementary symmetric polynomial
\begin{equation} \label{eq: def elementary symmetric polynomial}
   \syme_0(x_1, \ldots, x_n) =1, \qquad \syme_k(x_1, \ldots, x_n) := \sum_{1\leq j_1< \cdots < j_k\leq n} \prod_{\ell=1}^k x_{j_\ell}.
\end{equation}
Then, a direct computation shows that
\begin{equation} \label{eq. Meijer G diff operator expansion}
    \prod_{l=1}^m (\zeta \partial_\zeta + \nu_l) = \zeta^m \partial_\zeta^m
    + \sum_{j=0}^{m-1} \sum_{l=0}^{m-j} \syme_{l}(\vnu) S(m-l,j) \zeta^j \partial_\zeta^j,
\end{equation}
where $S(n,k)$ denotes the Stirling number of the second kind \cite[Section 26.8]{NIST}.

\subsection{Outline of the proof}\label{sec: outline of the proof}

We now outline the proofs of the main results, emphasizing how the differential equation in Proposition~\ref{Prop. diff eq kappa} leads to the limiting correlation functions.
Rather than directly analyzing the double sum defining the prekernel, we study its characterizing differential equation whose inhomogeneous terms involve only single sums.

We first consider Theorems~\ref{thm:real-line} and \ref{thm:complex-line}.
To this end, in the real regime, we modify the prekernel $\kappa_N$ by
\begin{equation*}
    \widehat{\kappa}_N(z,w) :=\frac{\kappa_N(z,w)}{f_N(z,w)},
\end{equation*}
where $f_N$ is specifically chosen to be a non-vanishing holomorphic function with respect to $z$. An analogous modification is introduced for the prekernels $\kappa_N^{a,b}$ in the complex regime.
Applying the change of variables \eqref{def-xi^rho,p_N(z)} to Proposition~\ref{Prop. diff eq kappa}, we obtain an $m$-th order differential equation of the form
\begin{equation}\label{eq: outline diff eq}
    \sum_{k=0}^m a_{N,k}(z,w) \, \partial_z^k\widehat{\kappa}_N(z,w) = \widehat{I}_N(z,w),
\end{equation}
where the coefficients $a_{N,k}(z,w)$ are explicit and $\widehat{I}_N(z,w)$ is obtained from the three inhomogeneous terms in Proposition~\ref{Prop. diff eq kappa}.

The next step is to analyze the large-$N$ asymptotics of the coefficients $a_{N,k}(z,w)$ and the inhomogeneous term $\widehat{I}_N(z,w)$ under the microscopic rescaling. The necessary estimates are established in Sections~\ref{sec:asymptotic} and \ref{sec:proofs}. In particular, we show that the higher-order differential terms become asymptotically negligible in the real regime, yielding a first-order differential equation. In the complex regime, all derivative terms disappear in the limit, leaving an equation that directly determines the limiting prekernel.

To justify the convergence of the modified prekernels $\widehat{\kappa}_N(z,w)$ and $\widehat{\kappa}_N^{a,b}(z,w)$, we establish their local uniform boundedness.
In the real regime, $\widehat{\kappa}_N(z,w)$ is holomorphic in both variables, so Montel's theorem yields locally uniformly convergent subsequences whose limits satisfy the limiting first-order differential equation; combined with the boundary condition $\widehat{\kappa}(z,z)=0$, uniqueness of the solution implies local uniform convergence of the full sequence. In the complex regime, holomorphicity in $z$ and Cauchy's integral estimates bound all $z$-derivatives locally uniformly, so the vanishing derivative terms in the differential equation directly yield local uniform convergence of $\widehat{\kappa}_N^{a,b}(z,w)$.

Finally, combining the convergence of the modified prekernels $\widehat{\kappa}_N(z,w)$ and $\widehat{\kappa}_N^{a,b}(z,w)$ with the asymptotics of the weight function $\omega_{\vnu}$ and the auxiliary factors $f_N(z,w)$, we obtain the limiting correlation functions. In the real regime, this yields the Pfaffian kernels stated in Theorem~\ref{thm:real-line}. In the complex regime, the diagonal entries of the $2\times2$ Pfaffian blocks vanish in the limit, and the resulting Pfaffian reduces to a determinant, leading to Theorem~\ref{thm:complex-line}.

\smallskip

For Theorem~\ref{thm:origin}, we also make use of the differential equation in Proposition~\ref{Prop. diff eq kappa}. In this regime, the induced parameters $\nu_\ell$ remain fixed, and the finite-$N$ prekernel converges directly to an infinite series. Unlike the regular scaling regimes, the limiting differential equation retains its full $m$-th order structure. By exploiting the differential equations satisfied by generalized hypergeometric functions, we construct an integral representation of the limiting prekernel and verify that it agrees with the limiting series. This establishes the parameter-dependent Pfaffian kernel in Theorem~\ref{thm:origin}.

\smallskip

We conclude by remarking that the differential-equation approach may also be useful for deriving higher-order asymptotic expansions of the prekernel. Beyond the leading order, the differential terms that become negligible in the above analysis would contribute to the equations governing the corrections. These contributions may depend on the parameters of the underlying model, suggesting that the subleading corrections need not be universal. Such parameter-dependent corrections have been derived, for instance, in the symplectic elliptic Ginibre ensemble \cite{BE23}. A systematic investigation of these corrections is beyond the scope of the present work.

\section{Asymptotic analysis }
\label{sec:asymptotic} 
This section is devoted to establishing the estimates needed for the asymptotic analysis of the differential equation obtained in Proposition~\ref{Prop. diff eq kappa}. There are two main difficulties to address. The first concerns the implicit characterization of the macroscopic density $\rho(p)$, which determines the microscopic scaling \eqref{eq. micro scale zeta eta z w}.
The second arises from the asymptotic analysis of the weight function $\omega_{\vnu}$ and the truncated sums appearing in the inhomogeneous terms of the differential equation. 

We first address the implicit characterization of the density. Recall that $\rho(p)$ in \eqref{eq: def rho macroscopic density} is expressed in terms of the root $\lambda_{\val}(p)$ of the degree-$m$ polynomial equation \eqref{eq:lambda-root}.
Since this root does not admit a convenient explicit expression for general $m$, we introduce an auxiliary function that allows us to analyze its behavior under microscopic rescaling.
More precisely, for $\vnu=(\nu_1,\ldots,\nu_m)$, let $\Upsilon_{\vnu}(\zeta,\eta)$ be the largest real root of
\begin{equation} \label{eq: def Upsilon nu}
    \prod_{\ell=1}^{m}
    \bigl(\Upsilon_{\vnu}(\zeta,\eta)+\nu_\ell\bigr)
    =\zeta\overline{\eta},
\end{equation}
whenever $\zeta\overline{\eta}\in(0,\infty)$. This root is uniquely determined and simple, and hence admits an analytic continuation to a neighborhood of the positive real axis in the variable $\zeta\overline{\eta}$.
In this way, $\Upsilon_{\vnu}(\zeta,\eta)$ extends to a function that is holomorphic in $\zeta$ and anti-holomorphic in $\eta$ in a neighborhood of the set 
\begin{equation*}
    \left\{(\zeta,\eta)\in\mathbb C^2: \zeta\overline{\eta}\in(0,\infty)\right\}.
\end{equation*}
In particular, for any fixed $p\neq0$ and $z,w$ in a compact subset of $\Cset$, the quantity $\Upsilon_{\vnu}(\xi_N^p(z), \xi_N^p(w) )$
is well-defined for all sufficiently large $N$. The asymptotic behavior of this root, which plays a central role in the subsequent analysis, is established in Section~\ref{sec: root asymp}.

We next turn to the weight function and the truncated sums. These objects also arise in the study of local eigenvalue correlations for products of induced complex Ginibre matrices; see, for instance, \cite{LW16}.
For our purposes, however, we require asymptotic estimates in a broader range of parameters and microscopic scaling regimes. An important feature of these estimates is that they can be expressed in terms of the auxiliary root $\Upsilon_{\vnu}$ introduced above.
To formulate the results, let $M\in\mathbb N\cup\{\infty\}$, where $M$ may depend on $N$. We are primarily interested in the cases where $M=\infty$ or $M$ grows linearly with $N$. Define the truncated sum
\begin{equation} \label{eq. def truncated sum T-M-a}
    T_{M,\vnu}(\zeta,\eta) := \sum_{k=0}^{M-1} \frac{(\zeta\overline{\eta})^k} {\prod_{\ell=1}^m \Gamma(\nu_\ell+k+1)},
\end{equation}
where the sum extends over all $k\geq0$ when $M=\infty$. Under the unfolding \eqref{def-xi^rho,p_N(z)}, we further write
\begin{equation} \label{eq. def truncated sum T-M-N-rho-p}
    T_{M,N,\vnu}^p(z,w) := T_{M,\vnu} \bigl(\xi_N^p(z),\xi_N^p(w)\bigr).
\end{equation}
When $M=N$, we use the abbreviated notation $$ T_{N,\vnu}^p(z,w) :=T_{N,N,\vnu}^p(z,w).$$
The required asymptotics of these sums and the weight function $\omega_{\vnu}$ are established in Section~\ref{sec: trunc sum and weight func}.

\subsection{Algebraic root and its asymptotics}
\label{sec: root asymp}

In this subsection, we study the algebraic root $\Upsilon_{\vnu}$ defined in \eqref{eq: def Upsilon nu}.
We derive algebraic identities relating its derivatives, and estimate their large-$N$ asymptotics under the microscopic rescaling \eqref{def-xi^rho,p_N(z)}.

In the sequel, we use the partial Bell polynomials defined for $1\leq k\leq n$ by
\begin{equation} \label{eq. def of Bell poly}
    B_{n,k}(x_1, \ldots, x_{n-k+1}) := n! \sum_{j_1, \ldots, j_{n-k+1}} \prod_{\ell=1}^{n-k+1} \frac{x_\ell^{j_\ell}}{(\ell!)^{j_\ell} j_\ell !},
\end{equation}
where the sum runs over all nonnegative integers $j_1, \ldots, j_{n-k+1}$ satisfying $\sum_\ell j_\ell =k$ and $\sum_\ell \ell \,j_\ell =n$ with conventions $B_{0,0}:=1$ and $B_{n,0}:=0$ for $n\geq 1$.
Then, the $n$-th complete Bell polynomial $B_{n}(x_1, \ldots, x_n)$ is defined by
\begin{equation}
    B_n(x_1, \ldots, x_n) := \sum_{k=0}^n B_{n,k}(x_1, \ldots, x_{n-k+1}).
\end{equation}
We recall the elementary symmetric polynomials $\syme_k$ from \eqref{eq: def elementary symmetric polynomial}.
For the rest of the manuscript, given $\vnu = (\nu_1, \ldots, \nu_m) \in \C^m$ and $x \in \C$, we denote by
\begin{equation*}
    \vnu + x := (\nu_1 +x, \ldots, \nu_m+x) \in \C^m.
\end{equation*}

\begin{lemma}
    Write $\Upsilon_{\vnu} := \Upsilon_{\vnu}(\zeta, \eta)$.
    Then, we have
    \begin{equation} \label{eq: elem sym poly partial-Upsilon}
        \syme_{m-1}(\Upsilon_{\vnu}+ \vnu) \,\partial_\zeta \Upsilon_{\vnu} - \overline{\eta} =0,
    \end{equation}
    and for every integer $j \geq 2$, that
    \begin{align} \label{eq. elem symm poly and Bell poly}
    \begin{split}
        \sum_{k=1}^{\min(j,m)} k! \, \syme_{m-k}(\Upsilon_{\vnu}+ \vnu) \, B_{j,k}\left(\partial_\zeta \Upsilon_{\vnu}, \ldots, \partial_\zeta^{j-k+1} \Upsilon_{\vnu}\right) = 0.
    \end{split}
    \end{align}
    Further assuming $\zeta, \eta \neq 0$, we have for every integer $j \geq 1$ that
    \begin{equation} \label{eq. power sum poly and Bell poly}
        \sum_{k=1}^j \sum_{l=1}^m \frac{(-1)^{k-1}(k-1)!}{(\Upsilon_{\vnu}+\nu_l)^k} B_{j,k}\left(\partial_\zeta \Upsilon_{\vnu}, \ldots, \partial_\zeta^{j-k+1}\Upsilon_{\vnu}\right) - (-1)^{j-1} \frac{(j-1)!}{\zeta^j} = 0.
    \end{equation}
\end{lemma}
\begin{proof}
    The first two assertions follow directly from Fa\`a di Bruno's formula
    \begin{align} \label{eq: Faa di Bruno}
        \frac{\dif^n}{\dif \zeta^n}f(g(\zeta)) = \sum_{k=0}^n f^{(k)}(g(\zeta))B_{n,k}(g'(\zeta),\dots,g^{(n-k+1)}(\zeta)),
    \end{align}
    applied to \eqref{eq: def Upsilon nu}.
    Taking the logarithm of \eqref{eq: def Upsilon nu}, we have
    \begin{equation*}
        \sum_{\ell=1}^m \log(\Upsilon_{\vnu}+\nu_\ell) - \log(\zeta \overline{\eta}) = 0.
    \end{equation*}
    Applying Fa\`a di Bruno's formula once more yields the last assertion.
\end{proof}

\begin{lemma} \label{lem: partial-Upsilon asymp}
    Suppose $p \in \C \setminus\{0\}$.
    Then, for every fixed integer $j\geq 0$, we have as $N\to\infty$ that
    \begin{equation} \label{eq: partial-Upsilon asymp}
        \partial_\zeta^j \Upsilon_{\vnu} \left(\zeta, \eta \right) \Big|_{\zeta= \xi_N^p(z), \eta = \xi_N^p(w)} = O(N^{1 - \frac{m}{2}j})
    \end{equation}
    locally uniformly for $z,w \in \Cset$.
\end{lemma}
\begin{proof}
    Let us define
    \begin{equation*}
        P_N(x) := \prod_{\ell=1}^m \Big( x + \frac{\nu_\ell}{N} \Big),
        \qquad
        P(x) := \prod_{\ell=1}^m (x+ \alpha_\ell).
    \end{equation*}
    Then, from \eqref{eq:lambda-root}, we have $P(\lambda_{\val}(p)) = |p|^2$.
    Since
    \begin{equation*}
        P'(\lambda_{\val}(p)) = |p|^2 \sum_{\ell=1}^m \frac{1}{\lambda_{\val}(p) + \alpha_\ell} >0,
        \qquad
        P_N'(x) - P'(x) = O\Big( \frac{1}{N} \Big),
    \end{equation*}
    the inverse function theorem implies that, for all sufficiently large $N$, the inverse $Q_N:=P_N^{-1}$ is a well-defined holomorphic function in a fixed neighborhood $U$ of $|p|^2$, whose derivatives of any fixed order are uniformly bounded in $U$.
    Moreover, by \eqref{eq: def Upsilon nu}, we have
    \begin{equation} \label{eq: Upsilon-vnu and Q-N}
        \Upsilon_{\vnu}(\zeta, \eta) = N Q_N\Big( \frac{\zeta \cconj \eta}{N^m} \Big),
        \qquad
        \partial_\zeta^j \Upsilon_{\vnu}(\zeta, \eta) = N  \Big( \frac{\cconj \eta}{N^m} \Big)^j Q_N^{(j)}\Big( \frac{\zeta \cconj \eta}{N^m} \Big)
    \end{equation}
    for any $\zeta \cconj \eta / N^m \in U$.
    Indeed, this is fulfilled for $\zeta= \xi_N^p(z)$ and $\eta= \xi_N^p(w)$ for any $z,w$ in a compact subset of $\C$. Since the derivatives of $Q_N$ are locally uniformly bounded, the conclusion follows.
\end{proof}

We next obtain a more precise expansion of $\Upsilon_{\vnu}$. Let us denote by
\begin{equation} \label{eq: def mu-p and sigma-val-p}
    \mu^p(z,w) := \cconj{p} \euler^{\iunit \theta} z + p \euler^{-\iunit \theta} \overline{w},
    \qquad
    \sigma_{\val}^p(k) := \sum_{\ell=1}^m \frac{1}{(\lambda_{\val}(p) + \alpha_\ell )^k}.
\end{equation}

\begin{lemma} \label{lem. lambda vs upsilon asymp}
    Suppose $p \in \C \setminus \{0\}$, and denote by $\zeta =\xi_N^p(z)$,  $\eta =\xi_N^p(w)$, $\xi_0 = \xi_N^p(0)$. Then as $N \to \infty$, we have
    \begin{equation} \label{eq. taylor expansion Upsilon-nu lambda-alpha}
        \Upsilon_{\vnu}(\xi_0, \xi_0) - N \lambda_{\val}(p) = - \rho(p) \, \vert{}p\vert{}^2 \sum_{\ell=1}^m \frac{\nu_\ell - N \alpha_\ell}{\lambda_{\val}(p) + \alpha_\ell} + O(N^{-1})
    \end{equation}
    and
    \begin{equation} \label{eq. taylor expansion of root Upsilon-vnu}
        \Upsilon_{\vnu}(\zeta,\eta) - \Upsilon_{\vnu}(\xi_0, \xi_0)
        = \sqrt{N\rho(p)} \, \mu^p(z,w)  - \frac{\mu^p(z,w)^2}{2 |p|^2}\Big( 1 -\frac{\sigma_{\val}^p(2)}{\sigma_{\val}^p(1)^2} \Big) + z \overline{w} + O(N^{-\frac{1}{2}}),
    \end{equation}
    locally uniformly for $z,w\in\C$, where $\mu^p$ and $\sigma_{\val}^p$ are given in \eqref{eq: def mu-p and sigma-val-p}.
\end{lemma}
\begin{proof}
    Recall the functions $P_N$ and $Q_N = P_N^{-1}$ from the proof of Lemma~\ref{lem: partial-Upsilon asymp}, and define
    \begin{equation*}
        L_N:=\Upsilon_{\vnu}(\xi_0, \xi_0) - N \lambda_{\val}(p).
    \end{equation*}
    Since $\nu_\ell - N \alpha_\ell = O(1)$, we deduce from \eqref{eq:lambda-root} and \eqref{eq: def Upsilon nu} that $L_N = O(1)$.
    Dividing \eqref{eq: def Upsilon nu} by \eqref{eq:lambda-root}, it follows from Taylor expansion that
    \begin{align*}
        0 &=\sum_{\ell=1}^m \log\Big(\frac{\Upsilon_{\vnu}(\xi_0,\xi_0) + \nu_\ell}{N(\lambda_{\val}(p) +  \alpha_\ell)}\Big)
        = \sum_{\ell=1}^m \log \Big( 1 + \frac{1}{N} \frac{L_N + \nu_\ell - N \alpha_\ell}{\lambda_{\val}(p) + \alpha_\ell} \Big)
        \\
        &= \frac{1}{N} \bigg( L_N \sigma_{\val}^p(1) + \sum_{\ell=1}^m \frac{\nu_\ell -N \alpha_\ell}{\lambda_{\val}(p)+\alpha_\ell} \bigg) + O (N^{-2}).
    \end{align*}
    Since $\sigma_{\val}^p(1) = \rho(p)^{-1} |p|^{-2}$, this proves \eqref{eq. taylor expansion Upsilon-nu lambda-alpha}.

    For the second assertion, we observe that
    \begin{equation*}
        \prod_{\ell=1}^m \Big(Q_N(x) + \frac{\nu_\ell}{N} \Big) = P_N(Q_N(x)) = x.
    \end{equation*}
    Differentiating it with respect to $x$, we obtain
    \begin{equation*}
        Q_N'(x) = \frac{1}{x \sum_{\ell=1}^m (Q_N(x) + \frac{\nu_\ell}{N})^{-1} },
        \qquad
        Q_N''(x) = - \frac{Q_N'(x)}{x} \bigg( 1 - \frac{\sum_{\ell=1}^m (Q_N(x) + \frac{\nu_\ell}{N})^{-2}}{(\sum_{\ell=1}^m (Q_N(x) + \frac{\nu_\ell}{N})^{-1})^2} \bigg).
    \end{equation*}
    Since $Q_N(|p|^2) = \lambda_{\val}(p) + O(N^{-1})$, we deduce that
    \begin{equation*}
        Q_N'(|p|^2) = \rho(p) + O(N^{-1}),
        \qquad
        Q_N''(|p|^2) = - \frac{\rho(p)}{|p|^2} \Big( 1 - \frac{\sigma_{\val}^p(2)}{\sigma_{\val}^p(1)^2} \Big) + O(N^{-1}).
    \end{equation*}
    On the other hand, from \eqref{eq. micro scale zeta eta z w}, we have
    \begin{equation*}
        \frac{\zeta\cconj \eta}{N^m} = |p|^2 +\frac{\mu^p(z,w)}{\sqrt{N\rho(p)}} +\frac{z\overline w}{N\rho(p)}.
    \end{equation*}
    Using \eqref{eq: Upsilon-vnu and Q-N} and applying Taylor's theorem for $Q_N$ around $|p|^2$, we obtain
    \begin{align*}
        \Upsilon_{\vnu}(\zeta,\eta) - \Upsilon_{\vnu}(\xi_0,\xi_0)
        = N Q_N'(|p|^2) \Big( \frac{\mu^p(z,w)}{\sqrt{N\rho(p)}} +\frac{z\overline w}{N\rho(p)} \Big) +\frac{Q_N''(|p|^2)}{2\rho(p)} \mu^p(z,w)^2 +O(N^{-1/2}).
    \end{align*}
    Substituting the above estimates for $Q_N'$ and $Q_N''$ yields \eqref{eq. taylor expansion of root Upsilon-vnu}.
    This finishes the proof.
\end{proof}

\subsection{Analysis of truncated sums and weight function}
\label{sec: trunc sum and weight func}
In this subsection, we establish the large-$N$ asymptotics of the truncated sum $T_{M,N,\vnu}^p(z,w)$ defined in \eqref{eq. def truncated sum T-M-N-rho-p} and the weight function $\omega_{\vnu}(\xi_N^p)$ defined in \eqref{eq. weight function omega m}.
These have also been studied in connection with the local eigenvalue correlations of products of induced complex Ginibre matrices \cite{LW16}. Our analysis builds on the multidimensional saddle-point method developed there, with the asymptotic formulas expressed in terms of the algebraic root $\Upsilon_{\vnu}$ introduced in \eqref{eq: def Upsilon nu}.

\begin{lemma} \label{lem: asymp Meijer-G}
    Suppose $p \in \C \setminus \{0\}$ and $\zeta = \xi_N^p(z)$.
    Then, as $N \to \infty$, we have
    \begin{equation}
    \begin{split} \label{eq. asymp Meijer G}
        G^{m,0}_{0,m}\Big( |\zeta|^2 \Big\vert \genfrac{}{}{0pt}{}{-}{\nu_1, \ldots, \nu_m} \Big) &=
        (2\pi)^{\frac{m-1}{2}} \bigg( \sum_{\ell=1}^m \frac{1}{\Upsilon_{\vnu}(\zeta, \zeta) + \nu_\ell} \bigg)^{-\frac{1}{2}} |\zeta|^{-1} 
        \\
        & \quad \times \exp\bigg[ - m \Upsilon_{\vnu}(\zeta, \zeta) - \sum_{\ell=1}^m \nu_\ell + \sum_{\ell=1}^m  \nu_\ell \log\big(\Upsilon_{\vnu}(\zeta, \zeta) + \nu_\ell \big)  \bigg] ( 1 + o(1) ),
    \end{split}
    \end{equation}
    locally uniformly for $z\in\C$.
\end{lemma}
\begin{proof}
    The multidimensional saddle-point analysis parallel to \cite[Lemma 4.1]{LW16} gives rise to the asymptotics for the Meijer $G$-function, as $N\to\infty$,
    \begin{equation} \label{eq. Meijer G ass from L4.1}
        G^{m,0}_{0,m}\Big( |\xi_N^p(z)|^2 \Big\vert \genfrac{}{}{0pt}{}{-}{\nu_1, \ldots, \nu_m} \Big) 
        = H_{N,\vnu}^p(z,z) (1 + o(1)),
    \end{equation}
    locally uniformly for $z\in\C$, where
    \begin{equation}
    \begin{split}  \label{eq. def of H-N-va-rho-p}
        H_{N,\vnu}^p(z,w) &:= \Big( \frac{2\pi}{N} \Big)^{\frac{m-1}{2}} \frac{1}{|p|}  \bigg( \sum_{\ell=1}^m \frac{N}{\Upsilon_{\vnu}^0+\nu_\ell} \bigg)^{-\frac{1}{2}} \Big( 1 + \frac{z\euler^{\iunit\theta}}{p\sqrt{N\rho}}  \Big)^{\nu_{\rm min}} \Big( 1 + \frac{\euler^{-\iunit\theta}\cconj{w}}{\cconj{p}\sqrt{N\rho}}   \Big)^{\nu_{\rm min}}
        \\
        \times& \exp\bigg[ - m \Upsilon_{\vnu}^0 - \sum_{\ell=1}^m \nu_\ell + \sum_{\ell=1}^{m}\nu_\ell \log (\Upsilon_{\vnu}^0 + \nu_\ell) - (\Upsilon_{\vnu}^0 + \nu_{\rm min}) \frac{p \euler^{-\iunit\theta} \overline{w} + \overline{p} \euler^{\iunit\theta} z}{|p|^2 \sqrt{N \rho}} \bigg]
        \\
        \times& \exp\bigg[ \frac{1}{2} \frac{(\Upsilon_{\vnu}^0 + \nu_{\rm min})}{ N |p|^4} \frac{\mu^p(z,w)^2}{ \rho} \Big( 1 - \Big( \sum_{\ell=1}^m \frac{\Upsilon_{\vnu}^0 + \nu_{\rm min}}{\Upsilon_{\vnu}^0 + \nu_\ell} \Big)^{-1} \Big) - \frac{z \cconj{w}}{|p|^2 N \rho} (\Upsilon_{\vnu}^0 + \nu_{\rm min}) \bigg],
        \end{split}
    \end{equation}
    with $\rho := \rho(p)$, $\nu_{\rm min} := \min\{\nu_1, \ldots, \nu_m\}$, $\Upsilon_{\vnu}^0 := \Upsilon_{\vnu}(\xi_N^p(0), \xi_N^p(0))$, and $\eta = \xi_N^p(w)$.
    After lengthy but straightforward computations using \eqref{eq. taylor expansion of root Upsilon-vnu} and Taylor expansion, it follows from \eqref{eq. def of H-N-va-rho-p} that
    \begin{equation} \label{eq. asymp H Na rho p simple ver}
    \begin{split}
        H_{N,\vnu}^p(z,w) &= (2\pi)^{\frac{m-1}{2}} \bigg( \sum_{l=1}^m \frac{1}{\Upsilon_{\vnu}(\zeta,\eta) + \nu_l} \bigg)^{-\frac{1}{2}} (\zeta \overline{\eta})^{-\frac{1}{2}} 
        \\
        & \quad \times \exp\bigg[ - m \Upsilon_{\vnu}(\zeta,\eta) - \sum_{l=1}^m \nu_l + \sum_{l=1}^m \nu_l \log(\Upsilon_{\vnu}(\zeta,\eta)+ \nu_l) \bigg] \Big( 1 + O(N^{-\frac{1}{2}}) \Big).
    \end{split}
    \end{equation}
    This is the desired result.
\end{proof}

We next turn to asymptotics of the truncated sum.
It is convenient to introduce the notation
\begin{equation} \label{eq: def U-N-vnu-p}
\begin{split}
    U_{N,\vnu}^p(z,w) &:= (2\pi)^{- \frac{m-1}{2}} \Big( \sum_{l=1}^m \frac{1}{\Upsilon_{\vnu}(\zeta, \eta) + \nu_l} \Big)^{-\frac{1}{2}} (\zeta \overline{\eta})^{-\frac{1}{2}}
    \\
    &\qquad \times \exp\bigg[ m \Upsilon_{\vnu}(\zeta,\eta) + \sum_{l=1}^m \nu_l - \sum_{l=1}^m \nu_l \log\big(\Upsilon_{\vnu}(\zeta,\eta) + \nu_l\big) \bigg],
\end{split}
\end{equation}
where $\zeta = \xi_N^p(z)$ and $\eta = \xi_N^p(w)$.

\begin{lemma} \label{lem: asymp T-M-N-vnu-p LW16 general}
    Suppose $p \in \C \setminus \{0\}$, and $M := M_N \in \mathbb{N} \cup \{\infty\}$ satisfies $M = cN + O(1)$ for fixed $c \in (0,\infty]$, as $N\to\infty$.
    Then, as $N\to\infty$,  we have
    \begin{equation}
    \begin{split} \label{eq. asymp T-N-a-rho-p}
        T_{M,N,\vnu}^p(z,w) &= U_{N,\vnu}^p(z,w) \big(h_{\val}^p(z,w) + o(1) \big),
    \end{split}
    \end{equation}
    locally uniformly for $z,w\in\C$, where $U_{N,\vnu}^p$ is defined in \eqref{eq: def U-N-vnu-p}, and
    \begin{equation} \label{eq. def of h-M-N-a-rho-p}
        h_{\val}^p(z,w) \equiv h_{\val}^p(z,w;c) := \begin{dcases}
            1, & \text{if} \quad |p| \in (r_{\rm in},  r_{c + \val}),
            \\
            \dfrac{1}{2} \erfc\Big( \frac{\mu^p(z,w)}{|p|\sqrt{2 }}\Big), & \text{if} \quad  |p| = r_{c + \val},
            \\
            \dfrac{1}{2} \erfc\Big( \frac{z + \overline{w}}{\sqrt{2 }}\Big), & \text{if} \quad |p| = r_{\rm in} > 0,
            \\
            0, & \text{if} \quad |p| \in (0, \infty) \setminus [r_{\rm in}, r_{c+ \val}].
        \end{dcases}
    \end{equation}
    Here, the function $\mu^p$ is defined in \eqref{eq: def mu-p and sigma-val-p}, and we set $r_{c+\val} := \infty$ if $c=\infty$ and, otherwise,
    \begin{equation}
        r_{c+\val} := \prod_{\ell=1}^m (c+ \alpha_\ell)^{1/2}.
    \end{equation}
\end{lemma}
\begin{proof}
    By symmetry in $\nu_1, \ldots, \nu_m$, we may assume without loss of generality that $\alpha_1 \geq \cdots \geq \alpha_m$ and, permuting indices with equal $\alpha_\ell$ for each $N$ if necessary, that $\nu_1 \geq \cdots \geq \nu_m$ for all sufficiently large $N$.
    
    We first prove the assertion assuming $c \in (0,\infty)$. In this case, the multi-dimensional saddle point analysis developed in \cite[Section 2]{LW16} applies.
    To avoid unnecessary repetition, we only give a brief sketch.
    As an analogue of \cite[Lemma 4.2]{LW16}, it follows from Hankel’s loop integral formula for the reciprocal Gamma function \cite[Eq.~(5.9.2)]{NIST} that the truncated sum $T_{M,N,\vnu}^p$ admits the $m$-fold contour representation
    \begin{equation*}
    \begin{aligned}
        T_{M,N,\vnu}^p(z,w)
        &=\frac{(2\pi \iunit)^{-m}}{N^{\nu_1 + \cdots+\nu_m}} \left(\frac{\xi_N^p(z)\cconj{\xi_N^p(w)}}{N^m}\right)^{-\nu_m}
        \\
        &\quad\times\int_{\gamma_1 \times \cdots \times \gamma_m}
        Q(t)\,\euler^{-\sqrt{N}\,F(t)}
        \frac{\euler^{-N P_1(t;\vnu)}- \euler^{-N P_2^{(M)}(t;\vnu)}}
        {t_1 \cdots t_m-1} \dif^m t,
    \end{aligned}
    \end{equation*}
    where, with $t=(t_1, \ldots, t_m)$, 
    \begin{align*}
        P_1(t;\vnu)&:= -\sum_{k=1}^{m-1}\Big(t_k-\frac{\nu_k}{N}\log t_k\Big)
        \ -\ |p|^2\,t_m +  \frac{\nu_m}{N}\log t_m,
        \\
        P_2^{(M)}(t;\vnu)&:=P_1(t; \vnu)+\frac{M}{N}\,\log (t_1 \cdots t_m) = P_1(t;\vnu+M),
    \end{align*}
    and
    \begin{equation*}
    F(t)=-\,\frac{\mu^p(z,w)}{\sqrt{\rho(p)}} t_m,
    \qquad
    Q(t)= \exp\Big(\frac{z\,\overline{w}}{\rho(p)}\,t_m\Big).
    \end{equation*}
    Here the paths $\gamma_k$ are constructed as follows. Let $\theta_0 \in (0,\pi/2)$ be fixed, and
    \begin{equation*}
        r_k := \lambda_{\val}(p)+\alpha_k \quad (k=1, \ldots, m-1)
        \qquad
        r_m:= 1/(r_1 \cdots r_{m-1}).
    \end{equation*}
    Then, the path $\gamma_k$ runs from $-\infty$ to $r_k \euler^{\iunit(\pi+\theta_0)}$ parallel to the real axis, traverses the circle of radius $r_k$ counterclockwise to $r_k \euler^{\iunit(\pi - \theta_0)}$, and returns to $-\infty$ parallel to the real axis.
    
    Using the assumption \eqref{eq: nu parameter assumption}, let us write $\nu_\ell=N\alpha_\ell +c_\ell$ for some $c_\ell = O(1)$. Then, we decompose
    \begin{equation*}
        P_1(t;\vnu)= P_1(t;N\val) + \frac{1}{N} \Big(c_m \log t_m + \sum_{k=1}^{m-1} c_k \log t_k\Big).
    \end{equation*}
    Notice that the function $P_1(t;N\val)$ is independent of $N$. 
    Therefore, after absorbing the subleading terms into the function $Q(t)$, the analysis in parallel to \cite[Sections 3, 4]{LW16} can be carried out.
    
    The truncation parameter $M$ enters solely through $P_2^{(M)}$, where it acts as a parameter shift $\vnu\rightarrow\vnu+M$ of $P_1$.
    Thus, up to shift of the parameter, similar analysis to that for $P_1$ applies. 
    As a consequence, we obtain
    \begin{equation*}
        T_{M,N,\vnu}^p(z,w) =  N^{-m} |p|^{-2} \bigg( \sum_{\ell=1}^m \frac{1}{N\lambda_{\val}(p)+N\alpha_\ell} \bigg)^{-1} \Big( H_{N,\vnu}^p(z,w)\Big)^{-1} (h_{\val}^p(z,w) + o(1)),
    \end{equation*} 
    where $H_{N,\vnu}^p(z,w)$ is defined in \eqref{eq. def of H-N-va-rho-p}.
    Taking into account \eqref{eq. asymp H Na rho p simple ver}, \eqref{eq: def U-N-vnu-p}, and Lemma~\ref{lem. lambda vs upsilon asymp}, we obtain the desired result for the case $c \in (0,\infty)$.

    We now consider the case $c=\infty$.
    Choose a constant $c_0\in(0,\infty)$ satisfying
    \begin{equation*}
        |p| < \prod_{\ell=1}^m \sqrt{\alpha_\ell + c_0}.
    \end{equation*}
    Then, we have a bound
    \begin{equation*}
        \big|T_{\infty,N,\vnu}^p(z,w) - T_{\lfloor c_0N \rfloor,N,\vnu}^p(z,w) \big| = \Bigg| \sum_{k= \lfloor c_0N \rfloor}^{\infty} \frac{(\xi_N^p(z) \overline{\xi_N^p(w)})^{k}}{\prod_{l=1}^m \Gamma(\nu_l + k + 1)} \Bigg|
        \leq  \sum_{k= \lfloor c_0 N \rfloor}^\infty \mathfrak{T}_k,
    \end{equation*}
    where
    \begin{equation} \label{eq: def mathfrak-T-k}
        \mathfrak{T}_k := \frac{|\zeta \eta|^k}{\prod_{\ell=1}^m \Gamma(\nu_\ell + k + 1)} \bigg|_{\zeta = \xi_N^p(z), \, \eta=\xi_N^p(w)}.
    \end{equation}
    By the choice of $c_0$, there is a constant $\delta=\delta(c_0) \in (0,1)$ such that $\mathfrak{T}_{k+1} / \mathfrak{T}_k < \delta $ for all $k\geq \lfloor c_0N \rfloor$. That is, $\mathfrak{T}_k$ exponentially decays in $k$ for all $k\geq \lfloor c_0N \rfloor$.
    Moreover, using Stirling's formula, a direct manipulation shows that
    \begin{equation*}
        \mathfrak{T}_{\lfloor c_0N \rfloor} = U_{N, \vnu}^p(z,w) \, O(\euler^{-\epsilon N})
    \end{equation*}
    for some $\epsilon>0$.
    Hence the conclusion follows from that for the case $c < \infty$.
\end{proof}

For the complex regime, we additionally need the following.

\begin{lemma} \label{lem. bound of T-N-va(z,conj z)}
    Suppose $p \in \C \setminus\R$ with $\zeta = \xi_N^p(z)$ and $\eta = \xi_N^p(w)$.
    Then as $N\to\infty$, we have
    \begin{equation}
        T_{N,\vnu}(\zeta, \overline{\eta}) = \begin{dcases}
            U_{N,\vnu}^p(z,w) \, O(N^{-1/2}), & \text{if} \quad |p| \in [r_{\rm in}, r_{\rm out}],
            \\
            U_{N,\vnu}^p(z,w) \, O(\euler^{-\epsilon N}), & \text{if} \quad  |p| \in (0, \infty) \setminus [r_{\rm in}, r_{\rm out}],
        \end{dcases}
    \end{equation}
    for some constant $\epsilon>0$, locally uniformly for $z,w\in\C$, where $U_{N,\vnu}^p$ is defined in \eqref{eq: def U-N-vnu-p}.
\end{lemma}
\begin{proof}
    Throughout the proof we identify $\zeta = \xi_N^p(z)$ and $\eta = \xi_N^p(w)$. From \eqref{eq. def truncated sum T-M-a}, we write
    \begin{equation*}
        T_{N,\vnu}(\zeta, \eta)= \sum_{k=0}^{N-1} \mathfrak{T}_k \euler^{\iunit k \phi_1},
        \qquad
        T_{N,\vnu}(\zeta, \overline{\eta}) = \sum_{k=0}^{N-1} \mathfrak{T}_k \euler^{\iunit k \phi_2},
    \end{equation*}
    where $\mathfrak{T}_k$ is defined in \eqref{eq: def mathfrak-T-k}, and $\phi_1, \phi_2 \in [0,2\pi)$ are defined by
    \begin{equation*}
        \phi_1 := \arg(\zeta \cconj{\eta}),
        \qquad
        \phi_2 := \arg(\zeta \eta).
    \end{equation*}
    Let us identify $\mathfrak{T}_k$ as a function of $k \in (-\nu_{\rm min} - 1, \infty)$, where $\nu_{\rm min}:= \min(\nu_1, \ldots, \nu_m)$.
    Since $\Gamma'(x)/\Gamma(x)$ is strictly increasing in $x>0$, the derivative
    \begin{equation} \label{eq: derivative log mathfrak-T}
        \frac{\dif}{\dif k} \log \mathfrak{T}_k = \log |\zeta \eta| - \sum_{\ell=1}^m \frac{\Gamma'(\nu_\ell + k +1)}{\Gamma(\nu_\ell + k +1)}
    \end{equation}
    is strictly decreasing in $k$.
    This implies $\mathfrak{T}_k$ is unimodal in $k \in (-\nu_{\rm min} - 1, \infty)$.

    Next, we consider large-$N$ asymptotics of $\mathfrak{T}_k$.
    For any $x>0$, the Gamma function satisfies \cite[Eq.(5.6.1)]{NIST}
    \begin{equation} \label{eq: stirling bound gamma}
        0<\log\Gamma(x) - \Big(x-\frac{1}{2}\Big) \log x + x - \frac{\log 2\pi}{2} < \frac{1}{12 x}.
    \end{equation}
    Let us denote by
    \begin{equation}
        k_\circ := \Upsilon_{\vnu}(|\zeta|, |\eta|),
        \qquad
        k_{\rm max} := \operatorname*{argmax}_{k \in \{0, \ldots, N-1\}} \mathfrak{T}_k,
        \qquad
        k_{\rm glob} := \operatorname*{argmax}_{k \in (-\nu_{\rm min} - 1, \infty)} \mathfrak{T}_k.
    \end{equation}
    The sum $T_{N,\vnu}(\zeta, \cconj{\eta})$ is bounded in terms of $\mathfrak{T}_{k_{\rm max}}$ as follows. Let $E_j := \sum_{k=0}^j \euler^{\iunit k \phi_2}$. Then,
    \begin{equation*}
        |E_j| = \Big| \frac{\euler^{\iunit (j+1) \phi_2}-1}{\euler^{\iunit \phi_2}-1} \Big| \leq \frac{2}{|\euler^{\iunit \phi_2}-1|} = \frac{1}{|\sin(\phi_2/2)|}
    \end{equation*}
    for any nonnegative integer $j$. Thus, for every integer $M \in \{1, \ldots, N\}$, we have
    \begin{align*}
        | T_{M,\vnu}(\zeta, \cconj{\eta}) |
        &= \Big| \mathfrak{T}_{M-1} E_{M-1} - \sum_{k=0}^{M-2} (\mathfrak{T}_{k+1} - \mathfrak{T}_k) E_k  \Big|
        \\
        &\leq \frac{1}{|\sin(\phi_2/2)|} \Big(  \mathfrak{T}_{M-1} + \sum_{k=0}^{M-2} |\mathfrak{T}_{k+1} - \mathfrak{T}_k| \Big) \leq \frac{3\mathfrak{T}_{k_{\rm max}}}{|\sin(\phi_2/2)|}.
    \end{align*}
    Since $p \in \C \setminus \R$, there is a constant $\delta >0$ such that $\phi_2 \in (\delta, 2\pi -\delta)$ for any sufficiently large $N$.
    Thus, as $N\to\infty$, we have
    \begin{equation} \label{eq: asymp T-n-vnu in terms of matfrak-T-kmax}
        \max_{1 \leq M \leq N} |T_{M,\vnu}(\zeta, \cconj{\eta})| = \mathfrak{T}_{k_{\rm max}} O(1). 
    \end{equation}

    It remains to estimate $\mathfrak{T}_{k_{\rm max}}$. Since $p \neq 0$, we have $\lambda_{\val}(p) + \alpha_\ell > 0$, and thus by Lemma~\ref{lem. lambda vs upsilon asymp},
    \begin{equation*}
        k_\circ + \nu_\ell = N(\lambda_{\val}(p) + \alpha_\ell) + O(\sqrt{N})
    \end{equation*}
    for all $\ell=1,\ldots,m$. Applying \eqref{eq: def Upsilon nu} for $k_\circ$ and \eqref{eq: stirling bound gamma}, we obtain
    \begin{align*}
        \log \mathfrak{T}_{k_\circ} & = \Phi(k_\circ) - \frac{m}{2} \log(2\pi) - \frac{1}{2} \log|\zeta \eta| + O(N^{-1}),
    \end{align*}
    where
    \begin{equation*}
        \Phi(k) := m k + \sum_{\ell=1}^m \nu_\ell - \sum_{\ell=1}^m \nu_\ell \log(k + \nu_\ell).
    \end{equation*}
    Comparing $|\zeta \eta| = |\zeta \cconj{\eta}|$ with $\zeta \cconj{\eta}$, we have $\Re[k_\circ - \Upsilon_{\vnu}(\zeta,\eta)] = O(1)$ and $\Im[k_\circ - \Upsilon_{\vnu}(\zeta,\eta)] = O(\sqrt{N})$ by Lemma~\ref{lem. lambda vs upsilon asymp}.
    Hence, expanding $\Phi(k)$ around $k=k_\circ$, where $\Phi'(k_\circ) = O(1)$ and $\Phi''(k_\circ) = O(N^{-1})$ are real, we deduce $$\Re[\Phi(k_\circ) - \Phi(\Upsilon_{\vnu}(\zeta,\eta))] = O(1).$$
    Therefore, from \eqref{eq: def U-N-vnu-p} and using $\sum_{\ell=1}^m (\Upsilon_{\vnu}(\zeta,\eta) + \nu_\ell)^{-1} = O(N^{-1})$, it follows that
    \begin{equation} \label{eq: mathfrak-T at k-circ}
        \mathfrak{T}_{k_\circ} = \big|U_{N,\vnu}^p(z,w)\big| \, O(N^{-1/2}).
    \end{equation}
    Furthermore, from \eqref{eq: derivative log mathfrak-T} and $\Gamma'(x)/\Gamma(x) = \log x + O(x^{-1})$ as $x \to \infty$ \cite[Eq.~(5.11.2)]{NIST}, we observe that
    \begin{equation*}
        k_{\rm glob} = k_\circ + O(1),
        \qquad
        \mathfrak{T}_{k_{\rm glob}} = \mathfrak{T}_{k_\circ}(1 + O(N^{-1})).
    \end{equation*}

    If $|p| \in [r_{\rm in}, r_{\rm out}]$, then
    $$
    \mathfrak{T}_{k_{\rm max}} \leq \mathfrak{T}_{k_{\rm glob}} = \big|U_{N,\vnu}^p(z,w)\big| \, O(N^{-1/2})
    $$
    by \eqref{eq: mathfrak-T at k-circ}.
    On the other hand, if $|p| > r_{\rm out}$, then $\lambda_{\val}(p) > 1$, so $k_{\rm glob} - N = N(\lambda_{\val}(p) - 1) + O(\sqrt{N})$ and $k_{\rm max} = N-1$ for sufficiently large $N$.
    Since $$\frac{\dif^2}{\dif k^2} \log \mathfrak{T}_k = -\sum_{\ell=1}^m (\log\Gamma)''(\nu_\ell + k + 1) \leq -c/N$$ on $k \in[N-1, k_{\rm glob}]$ for some constant $c>0$ \cite[Eq.~(5.15.8)]{NIST}, concavity yields
    $$
    \mathfrak{T}_{k_{\rm max}} = \mathfrak{T}_{N-1} = \mathfrak{T}_{k_{\rm glob}} O(\euler^{-\epsilon N}) = \big|U_{N,\vnu}^p(z,w)\big| \, O(\euler^{-\epsilon N})
    $$
    for some $\epsilon > 0$.
    Similarly, if $|p| \in (0, r_{\rm in})$, then $r_{\rm in} > 0$ so $\alpha_\ell > 0$ for all $\ell$, and $\lambda_{\val}(p) \in (-\alpha_{\rm min}, 0)$, hence $k_{\rm glob} = N\lambda_{\val}(p) + O(\sqrt{N}) < 0$ and $k_{\rm max} = 0$ for large $N$.
    Thus, using concavity of $\log \mathfrak{T}_k$ on $k \in [k_{\rm glob},0]$, we deduce that $\mathfrak{T}_{k_{\rm max}} = \mathfrak{T}_0 = \big|U_{N,\vnu}^p(z,w)\big| \, O(\euler^{-\epsilon N})$. Combining these bounds with \eqref{eq: asymp T-n-vnu in terms of matfrak-T-kmax} finishes the proof.
\end{proof}

\section{Scaling limits at the regular points}\label{sec:proofs}

This section is devoted to proving Theorems~\ref{thm:real-line} and \ref{thm:complex-line}.
In Sections~\ref{sec: auxiliary lemmas} and \ref{sec: inhomogeneous part}, we establish technical estimates required for their proof.
Theorems~\ref{thm:real-line} and \ref{thm:complex-line} are proved in Sections~\ref{sec: proof real regular} and \ref{sec: proof complex regular}, respectively.

Throughout this section, we denote by
\begin{equation} \label{eq: zeta, eta, eta-star notation}
    \zeta = \xi_N^p(z),
    \qquad
    \eta = \xi_N^p(w),
    \qquad
    \eta^* = \xi_N^p(\cconj{w}),
\end{equation}
where $\xi_N^p(\cdot)$ is defined in \eqref{def-xi^rho,p_N(z)}.

\subsection{Auxiliary lemmas} \label{sec: auxiliary lemmas}

We prove lemmas that will be used for the proof of Theorems \ref{thm:real-line} and \ref{thm:complex-line}.

\begin{lemma} \label{lem. bound of varkappa}
    Let $p \in \C \setminus \{0\}$, and let $\zeta$ and $\eta$ be as in \eqref{eq: zeta, eta, eta-star notation}.
    Then, for all sufficiently large $N$, we have
    \begin{equation} \label{eq. bound of varkappa}
        |\varkappa_N(\zeta,\eta)| \leq \frac{C}{\prod_{l=1}^m 4^{\nu_l}} U_{N,\vnu}^p(z,z) U_{N,\vnu}^p(w,w)
    \end{equation}
    for some constant $C>0$, locally uniformly for $z,w \in \C$.
    Furthermore, if $p \in \C \setminus \R$, then
    \begin{equation} \label{eq. bound of varkapp away from R}
        \max\Big\{|\varkappa_N(\zeta,\eta)|, |\varkappa_N(\zeta,\cconj\eta)|\Big\} \leq \frac{1}{\sqrt{N}} \frac{C}{\prod_{l=1}^m 4^{\nu_l}} U_{N,\vnu}^p(z,z) U_{N,\vnu}^p(w,w)
    \end{equation}
    for some constant $C>0$, locally uniformly for $z,w\in\C$. Here, $U_{N,\vnu}^p$ is defined in \eqref{eq: def U-N-vnu-p}.
\end{lemma}
\begin{proof}
    Using \eqref{eq. def of FN}, let us write $\varkappa_N(\zeta, \eta) = F_N(\zeta, \eta) - F_N(\eta, \zeta)$.
    Then, we observe that
    \begin{align} \label{eq. F-N(z,w) infinite sum}
        F_N(\zeta, \eta) &= c_m \sum_{k=0}^{N-1} \frac{\zeta^{2k+1}}{\prod_{l=1}^m \Gamma(\nu_l + k + \frac{3}{2})} \bigg[ \sum_{j=0}^\infty \frac{\eta^{2j}}{\prod_{l=1}^m \Gamma(\nu_l + j + 1)} - \sum_{j=k+1}^\infty \frac{\eta^{2j}}{\prod_{l=1}^m \Gamma(\nu_l + j + 1)}\bigg].
    \end{align}
    Thus, using $|T_{M,\vnu}(\zeta, \overline{\zeta})| \leq T_{M,\vnu}(\zeta, \zeta)$, we obtain
    \begin{align*}
        |F_N(\zeta,\eta)| \leq 2 c_m |\zeta| T_{N,\vnu + \frac{1}{2}}(\zeta, \zeta) T_{\infty, \vnu}(\eta, \eta) \leq 2 c_m |\zeta| T_{\infty,\vnu + \frac{1}{2}}(\zeta, \zeta) T_{\infty, \vnu}(\eta, \eta),
    \end{align*}
    where, for the second inequality, we used that $T_{M,\vnu}(\zeta, \zeta)$ is monotone increasing in $M$.
    By Lemma~\ref{lem: asymp T-M-N-vnu-p LW16 general} with $c = \infty$, since $|h_{\val}^p(z,z;\infty)| \leq 1$ for all $p \in \C \setminus \{0\}$, we have $T_{\infty,\vnu}(\eta,\eta) = T_{\infty,N,\vnu}^p(w,w) \leq C_1 U_{N,\vnu}^p(w,w)$ and $T_{\infty,\vnu+\frac{1}{2}}(\zeta,\zeta) = T_{\infty,N,\vnu+\frac{1}{2}}^p(z,z) \leq C_1 U_{N,\vnu+\frac{1}{2}}^p(z,z)$ for some constant $C_1 > 0$.
    Moreover, since $\Upsilon_{\vnu+\frac{1}{2}}(\zeta,\eta) = \Upsilon_{\vnu}(\zeta,\eta) - \frac{1}{2}$, it follows from \eqref{eq: def U-N-vnu-p} and \eqref{eq: def Upsilon nu} that
    \begin{equation} \label{eq: shift identity U-N-vnu}
        (\zeta \cconj{\eta})^{1/2} U_{N,\vnu+\frac{1}{2}}^p(z,w) = U_{N,\vnu}^p(z,w).
    \end{equation}
    Thus, there exists a constant $C>0$ such that
    \begin{equation*}
        |F_N(\zeta,\eta)| \leq \frac{C}{\prod_{\ell=1}^m 4^{\nu_\ell}} \, U_{N,\vnu}^p(z, z) U_{N, \vnu}^p(w,w).
    \end{equation*}
    A similar argument for $|F_N(\eta,\zeta)|$ finishes the proof of \eqref{eq. bound of varkappa}.

    We now further assume $p \in \C \setminus \R$.
    Denoting by
    \begin{equation*}
        b_k(\eta) := \sum_{j=0}^k \frac{\eta^{2j}}{\prod_{l=1}^m \Gamma(\nu_l + j + 1)} = T_{k+1, \vnu}(\eta, \cconj\eta),
        \qquad
        S_k(\zeta) := \sum_{r=0}^k \frac{\zeta^{2r+1}}{\prod_{l=1}^m \Gamma(\nu_l + r + \frac{3}{2})} = \zeta T_{k+1,\vnu+\frac{1}{2}}(\zeta,\cconj{\zeta}),
    \end{equation*}
    we use summation by parts in \eqref{eq. def of FN} to obtain
    \begin{equation*}
        F_N(\zeta,\eta) = c_m \bigg[ S_{N-1}(\zeta) b_{N-1}(\eta) - \sum_{k=0}^{N-2} S_k(\zeta) \big(b_{k+1}(\eta) - b_k(\eta)\big) \bigg].
    \end{equation*}
    Since $|b_{N-1}(\eta)| \leq T_{N,\vnu}(\eta,\eta)$ and $\sum_{k=0}^{N-2} |b_{k+1}(\eta) - b_k(\eta)| \leq T_{N,\vnu}(\eta,\eta) \leq T_{\infty,\vnu}(\eta,\eta)$, we obtain
    \begin{align*}
        |F_N(\zeta,\eta)| &\leq 2 c_m T_{\infty,\vnu}(\eta,\eta) \max_{1 \leq M \leq N} |\zeta T_{M,\vnu+\frac{1}{2}}(\zeta,\cconj{\zeta})|
        \leq  \frac{C}{\sqrt{N}\prod_{\ell=1}^m 4^{\nu_\ell}} U_{N,\vnu}^p(z,z) U_{N,\vnu}^p(w,w),
    \end{align*}
    where, for the second inequality, we used \eqref{eq: asymp T-n-vnu in terms of matfrak-T-kmax} and the bound on $\mathfrak{T}_{k_{\rm max}}$ in the proof of Lemma~\ref{lem. bound of T-N-va(z,conj z)}, together with \eqref{eq: shift identity U-N-vnu} and Lemma~\ref{lem: asymp T-M-N-vnu-p LW16 general}.
    Noting that $T_{\infty,\vnu}(\cconj{\eta},\cconj{\eta}) = T_{\infty,\vnu}(\eta,\eta)$ and $|T_{M,\vnu+\frac{1}{2}}(\cconj{\eta},\eta)| = |T_{M,\vnu+\frac{1}{2}}(\eta,\cconj{\eta})|$,
    the identical estimate holds for $|F_N(\eta,\zeta)|$, $|F_N(\zeta,\cconj{\eta})|$, and $|F_N(\cconj{\eta},\zeta)|$, proving \eqref{eq. bound of varkapp away from R}.
\end{proof}

Let $f_1(\zeta, \eta)$ and $f_2(\zeta, \eta)$ be functions defined by
\begin{align}
    f_1(\zeta, \eta) &:= \pi^{-\frac{m-1}{2}}  \bigg(  \sum_{l=1}^m \frac{1}{\Upsilon_{\vnu}(\zeta, \eta)+\nu_l} \bigg)^{\frac{1}{2}} (\zeta \overline{\eta})^{\frac{1}{2}}, \label{eq: def f1 auxiliary func}
    \\
    f_2(\zeta, \eta) &:= \exp\bigg( 2m \Upsilon_{\vnu}(\zeta, \eta) + \sum_{l=1}^m 2\nu_l -  \sum_{l=1}^m 2\nu_l \log(2\Upsilon_{\vnu}(\zeta, \eta) + 2\nu_l) \bigg), \label{eq: def f2 auxiliary func}
\end{align}
where $\Upsilon_{\vnu}$ is given in \eqref{eq: def Upsilon nu}. We note that these functions are holomorphic in $\zeta$, whereas antiholomorphic in $\eta$.

\begin{lemma} \label{lem. weight function and f1 f2}
    Let $p \in \C \setminus \{0\}$, and let $\zeta$ and $\eta$ be as in \eqref{eq: zeta, eta, eta-star notation}. Then, as $N\to\infty$, we have
    \begin{equation} 
        f_1(\zeta, \eta) f_2(\zeta, \eta) \sqrt{\omega_{\vnu}(\zeta) \omega_{\vnu}(\eta)} = c_N(z,w) \euler^{-|z|^2 - |w|^2 + 2z \overline{w} } (1+o(1)),
    \end{equation}
    locally uniformly for $z,w \in \C$, where $c_N(z,w)$ is a cocycle satisfying $c_N(\cconj{z}, \cconj{w}) = \cconj{c_N(z,w)}$ when $p \in \R \setminus \{0\}$.
\end{lemma}
\begin{proof}
    After straightforward manipulations using Lemma~\ref{lem. lambda vs upsilon asymp}, we deduce that
    \begin{equation*}
        f_1(\zeta,\eta) f_2(\zeta,\eta) = f_1(\xi_0,\xi_0) f_2(\xi_0,\xi_0) \, \exp\big[{\psi_N(z) + \cconj{\psi_N(w)} + 2z\cconj{w}}\big] \, \big(1 + O(N^{-\frac{1}{2}})\big),
    \end{equation*}
    where $\xi_0 := \xi_N^p(0)$, and denoting by $\Upsilon_{\vnu}^0:=\Upsilon_{\vnu}(\xi_0,\xi_0)$, 
    \begin{equation*}
        \psi_N(z) := \Big( \frac{2\euler^{\iunit\theta} z}{p \sqrt{N\rho(p)}} - \frac{\euler^{2\iunit\theta} z^2}{p^2 N \rho(p)} \Big) \Upsilon_{\vnu}^0  + \frac{\euler^{2\iunit\theta} z^2}{p^2 N \rho(p)} \Big( \sum_{\ell=1}^m \frac{1}{\Upsilon_{\vnu}^0 +\nu_\ell} \Big)^{-1}.
    \end{equation*}
    Substituting \eqref{eq. asymp Meijer G} into \eqref{eq. weight function omega m}, we obtain
    \begin{equation*}
        \omega_{\vnu}(\zeta) = \frac{1}{f_1(\zeta, \zeta) f_2(\zeta, \zeta)} (1 + o(1)) = \frac{\euler^{-2\Re \psi_N(z) - 2|z|^2}}{f_1(\xi_0, \xi_0) f_2(\xi_0, \xi_0)} (1 + o(1)).
    \end{equation*}
    Combining the above, the conclusion follows with $c_N(z,w) := \exp[\iunit(\Im\psi_N(z) - \Im\psi_N(w))]$.
\end{proof}

\begin{lemma}
    Let $p \in \C \setminus \{0\}$, and let $\zeta$ and $\eta$ be as in \eqref{eq: zeta, eta, eta-star notation}.
    Then, for any fixed integer $j\geq0$, as $N\to\infty$, we have
    \begin{align}
        \zeta^j \partial_\zeta^j f_1(\zeta,\eta) &=  f_1(\zeta, \eta) \, O(1), \label{eq. derivative f1 var param}
        \\
        \zeta^j \partial_\zeta^j f_2(\zeta,\eta) &=  f_2(\zeta, \eta) \big[ \big(2 \Upsilon_{\vnu}(\zeta,\eta)\big)^j + O(N^{j-1}) \big]  , \label{eq. derivative f2 var param}
    \end{align}
    locally uniformly for $z,w\in\C$.
\end{lemma}
\begin{proof}
    Throughout the proof we write $\Upsilon_{\vnu} := \Upsilon_{\vnu}(\zeta, \eta)$.
    Let us define
    \begin{align*}
        g_1(x) := \sum_{l=1}^m \frac{1}{x +  \nu_l},
        \qquad
        g_2(x) := 2 m x + \sum_{l=1}^m 2 \nu_l - \sum_{l=1}^m 2 \nu_l \log(2 x+ 2 \nu_l).
    \end{align*}
    Differentiating $f_1$ and $f_2$, it follows from the Leibniz rule and Fa\`a di Bruno's formula \eqref{eq: Faa di Bruno} that
    \begin{align}
        \partial_\zeta^{j} f_{1}(\zeta,\eta) &= (\pi)^{-\frac{m-1}{2}} \sum_{k=0}^j \binom{j}{k} \partial_\zeta^k\Big[ (g_1(\Upsilon_{\vnu}) )^{\frac{1}{2}} \Big] \partial_\zeta^{j-k} \Big[(\zeta \overline{\eta})^{\frac{1}{2}}\Big], \label{eq. partial Pa proof step 1}
        \\
        \partial_\zeta^j f_2(\zeta,\eta) &= \sum_{k=0}^j B_{j,k}(\partial_\zeta \Upsilon_{\vnu}, \ldots, \partial_\zeta^{j-k+1} \Upsilon_{\vnu}) B_k( g_2'(\Upsilon_{\vnu}), \ldots,  g_2^{(k)}(\Upsilon_{\vnu})) f_2(\zeta,\eta).
        \label{eq. partial f2 Bell partial g2}
    \end{align}
    For \eqref{eq. partial Pa proof step 1}, we again use Fa\`a di Bruno's formula, and obtain
    \begin{align*}
        \partial_\zeta^k\Big[ (g_1(\Upsilon_{\vnu}) )^{\frac{1}{2}} \Big] = \sum_{s=0}^k B_{k,s}(\partial_\zeta\Upsilon_{\vnu}, \ldots, \partial_\zeta^{k-s+1}\Upsilon_{\vnu}) \sum_{t=0}^s B_{s,t}(g_1'(\Upsilon_{\vnu}), \ldots, g_1^{(s-t+1)}(\Upsilon_{\vnu})) \frac{\Gamma(\frac{3}{2})}{\Gamma(\frac{3}{2} - t)} [ g_1(\Upsilon_{\vnu}) ]^{\frac{1}{2} - t}.
    \end{align*}
    By direct computations, we bound
    \begin{equation*}
        g_1^{(t)}(\Upsilon_{\vnu}) = (-1)^t t! \sum_{l=1}^m \frac{1}{(\Upsilon_{\vnu}+\nu_l)^{t+1}} = O(N^{-t-1}).
    \end{equation*}
    Moreover, by Lemma~\ref{lem: partial-Upsilon asymp}, we have
    \begin{align} \label{eq. Bell partial lambda asymp}
        B_{j+1,k}(\partial_\zeta \Upsilon_{\vnu}, \ldots, \partial_\zeta^{j-k+2}\Upsilon_{\vnu})
        = (j+1)! \sum_{r_1, \ldots, r_{j-k+2}} \prod_{\ell=1}^{j-k+2} \frac{(\partial_\zeta^{\ell}\Upsilon_{\vnu})^{r_\ell}}{(\ell!)^{r_\ell}r_\ell!}
        = O(N^{k - \frac{m}{2}(j+1)}), 
    \end{align}
    where the summation runs over nonnegative integers satisfying $\sum_{\ell=1}^{j-k+2} r_\ell = k$ and $\sum_{\ell=1}^{j-k+2} \ell \,r_\ell = j+1$.
    Hence, we have
    \begin{align*}
        \partial_\zeta^k\Big[ (g_1(\Upsilon_{\vnu}) )^{\frac{1}{2}} \Big] &= \sum_{s=0}^k O(N^{s-\frac{m}{2}k}) \sum_{t=0}^s O(N^{-s-t}) O(N^{t-\frac{1}{2}})
        = \zeta^{-k} O(N^{ - \frac{1}{2}}).
    \end{align*}
    Putting all of the above in \eqref{eq. partial Pa proof step 1}, we have
    \begin{align*}
        \partial_\zeta^j f_1(\zeta,\eta) &= (\pi)^{-\frac{m-1}{2}} \sum_{k=0}^j \binom{j}{k} \zeta^{-k} O(N^{-\frac{1}{2}}) \frac{\Gamma(\frac{3}{2})}{\Gamma(\frac{3}{2}-j+k)}  \zeta^{-j+k}(\zeta \overline{\eta})^{\frac{1}{2}}
        = \zeta^{-j} f_{1}(\zeta,\eta) \,O(1),
    \end{align*}
    where we used $f_{1}(\zeta, \eta) = O(N^{\frac{m-1}{2}})$ for the last equality. This implies \eqref{eq. derivative f1 var param}.
    
    We estimate $\partial_\zeta^j f_2$ similarly. It follows from direct computations that
    \begin{equation*}
        g_2'(x) = 2m -  \sum_{l=1}^m \frac{2\nu_l}{x+ \nu_l},
        \qquad
        g_2^{(k)}(x) = (-1)^k 2 (k-1)! \sum_{l=1}^m \frac{\nu_l}{(x+ \nu_l)^k} \quad \text{for} \quad k \geq 2.
    \end{equation*}
    Since $\Upsilon_{\vnu} = O(N)$ and $g_2^{(k)}(\Upsilon_{\vnu}) = O(N^{1-k})$, we have
    \begin{equation*}
        B_k( g_2'(\Upsilon_{\vnu}), \ldots,  g_2^{(k)}(\Upsilon_{\vnu})) = g_2'(\Upsilon_{\vnu})^k  + \sum_{\ell=1}^{k-1} O(N^{\ell-k})
        = g_2'(\Upsilon_{\vnu})^k+ O(N^{-1}).
    \end{equation*}
    Substituting it back into \eqref{eq. partial f2 Bell partial g2}, together with \eqref{eq. Bell partial lambda asymp}, we obtain
    \begin{align*}
        \partial_\zeta^j f_{2}(\zeta,\eta)
        &= B_{j,j}(\partial_\zeta\Upsilon_{\vnu}) \big( 1+O(N^{-1}) \big)  \big(g_2'(\Upsilon_{\vnu})^j  + O(N^{-1})\big) \, f_{2}(\zeta,\eta) 
        \\
        &= (\partial_\zeta \Upsilon_{\vnu})^j  \big(  g_2'(\Upsilon_{\vnu})^j +  O(N^{-1}) \big) f_{2}(\zeta,\eta) .
    \end{align*}
    In particular, from \eqref{eq. power sum poly and Bell poly}, we have
    \begin{equation*}
        g_2'(\Upsilon_{\vnu}) \partial_\zeta\Upsilon_{\vnu} = \sum_{l=1}^m \frac{2\Upsilon_{\vnu}}{\Upsilon_{\vnu}+\nu_l}  \partial_\zeta \Upsilon_{\vnu} = 2\Upsilon_{\vnu} \frac{1}{\zeta}.
    \end{equation*}
    Finally, using $\partial_\zeta \Upsilon_{\vnu} = \zeta^{-1} O(N)$ which follows from 
    \eqref{eq: elem sym poly partial-Upsilon}, we obtain the desired result.
\end{proof}

\subsection{Asymptotic analysis of the inhomogeneous part} \label{sec: inhomogeneous part}
In this section, we estimate the inhomogeneous terms in Proposition~\ref{Prop. diff eq kappa}.
For $j=1,2,3$, we define
\begin{align} \label{eq. def widehat r-N-j}
    \widehat{I}_{N,j}(z,w) := \frac{ I_{N,j}(\zeta,\cconj{\eta^*}) }{ 2^{m-1} N^{1-m} \rho(p) f_1(\zeta,\eta^*) f_2(\zeta,\eta^*) },
    \qquad
    \widetilde{I}_{N,j}(z,w) := \frac{ I_{N,j}(\zeta,\eta) }{ 2^{m-1} N^{1-m} \rho(p) f_1(\zeta,\eta) f_2(\zeta,\eta) },
\end{align}
where $\zeta$, $\eta$ and $\eta^*$ are given in \eqref{eq: zeta, eta, eta-star notation}.
Indeed, these functions are holomorphic in $z$.

\begin{lemma} \label{lem. asymp widehat r-N-1}
    If $p \in \C \setminus \{0\}$, then, as $N\to\infty$,
    \begin{align} \label{eq. asymp r N1}
        \widehat{I}_{N,1}(z,w) =\begin{dcases}
            \frac{4}{\pi} + o(1), & \text{if} \quad |p| \in (r_{\rm in}, r_{\rm out}),
            \\
            \frac{2}{\pi}  \erfc( z+w ) + o(1), & \text{if} \quad |p| = r_{\rm out} \quad \text{or} \quad |p| = r_{\rm in}>0,
            \\
            o(1), & \text{if} \quad |p| \in (0, \infty) \setminus [r_{\rm in}, r_{\rm out}],
        \end{dcases}
    \end{align}
    locally uniformly for $z, w \in \C$.
    Furthermore, if $p \in \C \setminus \R$, then
    \begin{equation}
        \label{eq. asymp I tilde N1 complex}
        \widetilde{I}_{N,1}(z,w)  = O(N^{-1/2}),
    \end{equation}
    locally uniformly for $z, w \in \C$.
\end{lemma}
\begin{proof}
    By the definitions \eqref{def. inhom part I_1} and \eqref{eq. def truncated sum T-M-a}, for general $\zeta$ and $\eta$, we have
    \begin{equation}
        \label{eq. I_N1 and T_2n,2v}
        I_{N,1}(\zeta,\eta) = \frac{2^{2m}}{\pi} T_{2N, 2\vnu}(2^{m/2}\zeta, 2^{m/2} \overline{\eta}).
    \end{equation}
    In particular, for $\zeta$ and $\eta^*$ as in \eqref{eq: zeta, eta, eta-star notation}, since $\xi_{2N}^p(\sqrt{2} z) = 2^\frac{m}{2} \xi_N^p(z)$, we have
    \begin{align*}
        I_{N,1}(\zeta,\cconj{\eta^*}) = \frac{2^{2m}}{\pi} T_{2N, 2\vnu}^p(\sqrt{2} z, \sqrt{2}\cconj{w}).
    \end{align*}
    Then, it follows from \eqref{eq. asymp T-N-a-rho-p} that
    \begin{align*}
        I_{N,1}(\zeta,\cconj{\eta^*}) = \frac{2^{2m}}{\pi} \bigg( \sum_{l=1}^m \frac{1}{2\Upsilon_{\vnu}(\zeta,\eta^*) + 2\nu_l} \bigg)^{-1} (2^m \zeta \cconj{\eta^*})^{-1} f_1(\zeta, \eta^*) f_2(\zeta, \eta^*) \Big( h_{\val}^p(\sqrt{2}z, \sqrt{2}\cconj{w}) + o(1) \Big).
    \end{align*}
    Dividing by the normalization factor in \eqref{eq. def widehat r-N-j}, we obtain
    \begin{align*}
        \widehat{I}_{N,1}(z,w) = \frac{4}{\pi} h_{\val}^p(\sqrt{2} z, \sqrt{2} \cconj{w}) + o(1),
    \end{align*}
    and the assertion \eqref{eq. asymp r N1} follows from \eqref{eq. def of h-M-N-a-rho-p}. 

    Let us assume $p \in \C \setminus \R$.
    Applying Lemma~\ref{lem. bound of T-N-va(z,conj z)} to \eqref{eq. I_N1 and T_2n,2v}, we obtain
    \begin{equation*}
        \big| I_{N,1}(\zeta,\eta) \big| = \big| U_{2N,2\vnu}^p(\sqrt{2}z, \sqrt{2}w) \big| \, O(N^{-1/2}).
    \end{equation*}
    Then, comparing \eqref{eq: def U-N-vnu-p} with \eqref{eq: def f1 auxiliary func} and \eqref{eq: def f2 auxiliary func} using $\Upsilon_{2\vnu}(2^{m/2}\zeta, 2^{m/2}\eta) = 2\Upsilon_{\vnu}(\zeta,\eta)$, the conclusion \eqref{eq. asymp I tilde N1 complex} follows from \eqref{eq. def widehat r-N-j}.
\end{proof}

\begin{lemma} \label{lem. asymp widehat r-N-2}
    If $p \in \Rset \setminus\{0\}$, then, as $N\to \infty$,
    \begin{align} \label{eq. asymp r-N-2}
        \widehat{I}_{N,2}(z,w) &= \begin{dcases}
            O(\euler^{-\epsilon N}), &\quad \text{if} \quad |p| \neq r_{\rm out},
            \\
            - \frac{\sqrt{2}}{\pi} \euler^{(z-w)^2-2 z^2} \erfc(\sqrt{2} w) + o(1), &\quad \text{if} \quad |p| = r_{\rm out},
        \end{dcases}
    \end{align}
    for some constant $\epsilon>0$, locally uniformly for $z, w \in \C$.
    Furthermore, if $p \in \C \setminus \R$, then
    \begin{equation}
        \label{eq. asymp I-N-2 complex}
        \widehat{I}_{N,2}(z,w) = O(N^{-1/2}),
        \qquad
        \widetilde{I}_{N,2}(z,w) = O(N^{-1/2}),
    \end{equation}
    locally uniformly for $z, w \in \C$.
\end{lemma}
\begin{proof}
    Throughout the proof, we always use \eqref{eq: zeta, eta, eta-star notation}, and identify $\xi_0 = \xi_N^p(0)$ and
    \begin{equation} \label{eq: Upsilon-vnu-zw notations}
        \Upsilon_{\vnu}^{z,w} := \Upsilon_{\vnu}(\zeta, \eta^*),
        \qquad
        \Upsilon_{\vnu}^w := \Upsilon_{\vnu}(\eta, \eta^*),
        \qquad
        \Upsilon_{\vnu}^0 := \Upsilon_{\vnu}(\xi_0, \xi_0).
    \end{equation}
    
    Let us first consider the case $p \in \R \setminus \{0\}$. Since $p = \cconj{p}$,  using \eqref{def. inhom part I_2} and \eqref{eq. def truncated sum T-M-N-rho-p}, we identify
    \begin{align*}
        I_{N,2}(\zeta,\cconj{\eta^*}) = I_{N,2}(\zeta,\eta) = - \frac{ 2^{2m} \pi^{\frac{m}{2}-1} \zeta^{2N} }{\prod_{l=1}^m 4^{\nu_l} \Gamma(\nu_l + N + 1/2) }   T_{N,\vnu}^p(w, \cconj{w}).
    \end{align*}
    The large-$N$ asymptotic of the prefactor can be computed using Stirling's formula, which yields
    \begin{align*}
    \begin{split}
        \frac{ 2^{2m} \pi^{\frac{m}{2}-1} \zeta^{2N} }{\prod_{l=1}^m 4^{\nu_l} \Gamma(\nu_l + N + 1/2) } &= 2^{\frac{3m}{2} - 2\sum_{l=1}^m \nu_l} \pi^{-1} \exp\bigg[ N \log (N^m p^2) - \sum_{l=1}^m (\nu_l + N) \log(\nu_l + N) \bigg]
        \\
        &\quad \times \exp\bigg[ \sum_{l=1}^m (\nu_l + N) + 2 N \frac{\euler^{\iunit\theta} z}{p \sqrt{N\rho}} - N \frac{\euler^{2\iunit\theta}z^2}{p^2 N \rho} + O(N^{-\frac{1}{2}}) \bigg].
    \end{split}
    \end{align*}
    Here and for the rest of the proof, we write $\rho := \rho(p)$ for brevity.
    Hence, combined with \eqref{eq. asymp T-N-a-rho-p}, we obtain
    \begin{align*}
    \begin{split} 
        I_{N,2}(\zeta,\eta) &= -2^{\frac{3m-1}{2}} \pi^{-1} (2\pi)^{-\frac{m-1}{2}} \bigg(  \sum_{l=1}^m \frac{1}{2 \Upsilon_{\vnu}^w + 2\nu_l} \bigg)^{-\frac{1}{2}} N^{- \frac{m}{2}} |p|^{-1} \\
        & \quad \times\exp\bigg[ N \log (N^mp^2) - \sum_{l=1}^m (\nu_l + N) \log(\nu_l + N) - \sum_{l=1}^m \nu_l \log(\Upsilon_{\vnu}^w + \nu_l) \bigg]
        \\
        &\quad \times \exp\bigg[ - 2 \log 2 \sum_{l=1}^m \nu_l + m (N + \Upsilon_{\vnu}^w) +  2\sum_{l=1}^m \nu_l \bigg]
        \\
        &\quad \times \exp\bigg[ 2 N \frac{\euler^{\iunit\theta}z}{p \sqrt{N\rho}} - N \frac{\euler^{2\iunit\theta}z^2}{p^2 N \rho} + O(N^{-\frac{1}{2}}) \bigg] \Big( h_{\val}^p(w, \cconj{w}) + o(1) \Big).
    \end{split}
    \end{align*}
    Using the identity $\prod_{\ell=1}^m (\Upsilon_{\vnu}^0 + \nu_\ell) = N^m p^2$ and rearranging terms, we obtain
    \begin{align}
    \begin{split} \label{eq. asymp widehat r-N-2 step1}
        \widehat{I}_{N,2}(z,w) &= - \frac{2 \sqrt{2}}{\pi} \exp\left[ \sum_{\ell=1}^m S_\ell +  2N \frac{\euler^{\iunit\theta} z}{p \sqrt{N\rho}} - N \frac{\euler^{2\iunit\theta}z^2}{p^2 N\rho}  + O(N^{-\frac{1}{2}}) \right] \Big( h_{\val}^p(w, \cconj{w}) + o(1) \Big),
    \end{split}
    \end{align}
    where
    \begin{align*}
        S_\ell := N \log \Big( \frac{\Upsilon_{\vnu}^0 + \nu_\ell}{N+\nu_\ell} \Big) - \nu_\ell \log \Big( \frac{\Upsilon_{\vnu}^w + \nu_\ell}{N+\nu_\ell} \Big) + 2 \nu_\ell \log \Big( \frac{\Upsilon_{\vnu}^{z,w} + \nu_\ell}{N+\nu_\ell} \Big) + N +\Upsilon_{\vnu}^w - 2 \Upsilon_{\vnu}^{z,w}.
    \end{align*}
    
    We further assume $|p| \neq r_{\rm out}$.
    In this case, we observe $\Upsilon_{\vnu}^0 /N \to \lambda_{\val}(p) \neq 1$.
    Moreover, by Lemma~\ref{lem. lambda vs upsilon asymp}, we have $\Upsilon_{\vnu}^0 = \Upsilon_{\vnu}^w + O(\sqrt{N}) = \Upsilon_{\vnu}^{z,w} + O(\sqrt{N})$.
    For any $\delta>0$, there exists $\epsilon_\delta>0$ such that $\log x < x-(1+\epsilon_\delta)$ for all $|x-1| >\delta$.
    Therefore, choosing $\delta>0$ sufficiently small, we have
    \begin{equation*}
        \Re S_\ell = (N+\nu_\ell) \log \Big( \frac{\Upsilon_{\vnu}^0 + \nu_\ell}{N+\nu_\ell} \Big) + N - \Upsilon_{\vnu}^0 + O(\sqrt{N}) < - \epsilon_\delta (N+\nu_\ell) + O(\sqrt{N}).
    \end{equation*}
    Substituting this bound into \eqref{eq. asymp widehat r-N-2 step1}, for any $\epsilon \in (0, \epsilon_\delta)$, we obtain
    \begin{equation*}
        \widehat{I}_{N,2}(z,w) = O(\euler^{-\epsilon N}).
    \end{equation*}

    Let us consider the case $p = \pm r_{\rm out}$.
    In this case, the exponent in \eqref{eq. asymp widehat r-N-2 step1} converges to a nontrivial function in the large-$N$ limit.
    To see this, we first use Taylor expansion for the logarithm functions in $S_\ell$, which yields
    \begin{align}
    \begin{split} \label{eq: S-ell taylor expanded}
         S_\ell &= N \log \Big( 1+ \frac{\Upsilon_{\vnu}^0 - N}{N+\nu_\ell} \Big) - \nu_\ell \log \Big( 1+ \frac{\Upsilon_{\vnu}^w - N}{N+\nu_\ell} \Big) + 2 \nu_\ell \log \Big( 1+ \frac{\Upsilon_{\vnu}^{z,w} -N}{N+\nu_\ell} \Big) + N +\Upsilon_{\vnu}^w - 2 \Upsilon_{\vnu}^{z,w}
         \\
         &= \frac{N (\Upsilon_{\vnu}^0 + \Upsilon_{\vnu}^w - 2\Upsilon_{\vnu}^{z,w})}{N+\nu_\ell} - \frac{ N (\Upsilon_{\vnu}^0-N)^2 - \nu_\ell (\Upsilon_{\vnu}^w-N)^2 + 2 \nu_\ell (\Upsilon_{\vnu}^{z,w}-N)^2}{2(N+\nu_\ell)^2} + O(N^{-\frac{1}{2}}).
     \end{split}
    \end{align}
    Since $N \lambda_{\val}(p) = N$, we utilise Lemma~\ref{lem. lambda vs upsilon asymp} to further expand \eqref{eq: S-ell taylor expanded}.
    After straightforward computations using \eqref{eq: def rho macroscopic density}, their summation simplifies to
    \begin{align*}
        \sum_{\ell=1}^m S_\ell  &=  (z-w)^2-2z^2 -  2N \frac{\euler^{\iunit\theta} z}{p \sqrt{N\rho}} + N \frac{\euler^{2\iunit\theta}z^2}{p^2 N\rho} + O(N^{-\frac{1}{2}}).
    \end{align*}
    Substituting it into \eqref{eq. asymp widehat r-N-2 step1}, the desired result follows from \eqref{eq. def of h-M-N-a-rho-p}.

    Finally, we consider the case $p\in\Cset\setminus\Rset$.
    The asymptotic formula \eqref{eq. asymp I-N-2 complex} can be proved similarly as in Lemma~\ref{lem. asymp widehat r-N-1}.
    More precisely, by \eqref{def. inhom part I_2} and \eqref{eq. def truncated sum T-M-a} together with Lemma~\ref{lem. bound of T-N-va(z,conj z)}, we have
    \begin{align*}
        I_{N,2}(\zeta,\eta) = - \frac{ 2^{2m} \pi^{\frac{m}{2}-1} \zeta^{2N} }{\prod_{l=1}^m 4^{\nu_l} \Gamma(\nu_l + N + 1/2) }   T_{N,\vnu}(\eta, \cconj{\eta}) = - \frac{ 2^{2m} \pi^{\frac{m}{2}-1} \zeta^{2N} }{\prod_{l=1}^m 4^{\nu_l} \Gamma(\nu_l + N + 1/2) }   U_{N,\vnu}^p(w,w) O(N^{-\frac{1}{2}})
    \end{align*}
    and an analogous bound holds for $I_{N,2}(\zeta,\cconj{\eta^*})$ with $w$ replaced by $\cconj{w}$.
    After similar computations as done in the real case for the real part of the exponent, with the additional factor $O(N^{-\frac{1}{2}})$, we conclude \eqref{eq. asymp I-N-2 complex}.
    This finishes the proof.
\end{proof}

\begin{lemma} \label{lem. asymp widehat r-N-3}
    If $p \in \Rset \setminus \{0\}$, then as $N \to \infty$, we have
    \begin{align} \label{eq. asymp r-N-3}
        \widehat{I}_{N,3}(z,w) &= \begin{dcases}
            O(\euler^{-\epsilon N}), &\quad \text{if} \quad |p| \neq r_{\rm in},
            \\
            - \frac{\sqrt{2}}{\pi} \euler^{(z-w)^2-2 z^2} \erfc(\sqrt{2}w) + o(1), &\quad \text{if} \quad |p| = r_{\rm in} >0,
        \end{dcases}
    \end{align}
    for some constant $\epsilon>0$, locally uniformly for $z, w \in \C$.
    Furthermore, if $p \in \C \setminus \R$, then we have
    \begin{equation}
        \widehat{I}_{N,3}(z,w) = O(N^{-\frac{1}{2}}), \qquad \text{and} \qquad \widetilde{I}_{N,3}(z,w) = O(N^{-\frac{1}{2}}),
    \end{equation}
    locally uniformly for $z, w \in \C$.
\end{lemma}
\begin{proof}
    The proof is similar to that of Lemma~\ref{lem. asymp widehat r-N-3} after some modifications.
    By \eqref{def. inhom part I_3} and \eqref{eq. def truncated sum T-M-a}, for any $\zeta, \eta \in \C$ with $\zeta \neq 0$, we have
    \begin{equation*}
        I_{N,3}(\zeta, \eta) = - \frac{2^{2m} \pi^{\frac{m}{2}-1}}{\prod_{\ell=1}^m 4^{\nu_\ell} \Gamma(\nu_\ell)} \frac{\eta}{\zeta} T_{N,\vnu+\frac{1}{2}}(\eta, \cconj{\eta}).
    \end{equation*}
    If $\nu_\ell = 0$ for some $\ell$, then $I_{N,3} \equiv 0$ by $1/\Gamma(0) = 0$ while $r_{\rm in} = 0 \neq |p|$, so both assertions hold trivially.
    Hence, we may assume $\nu_\ell \neq 0$ for all $\ell = 1, \ldots, m$.

    If $r_{\rm in}>0$, then $\alpha_\ell >0$ for all $\ell =1, \ldots, m$. Thus, similar arguments to Lemma~\ref{lem. asymp widehat r-N-2} (bounding $T_{N,\vnu+\frac{1}{2}}$ directly via a geometric series when $|p| < r_{\rm in}$) give rise to the desired assertions.
    On the other hand, if $r_{\rm in}=0$, then we have $\alpha_{\ell} =0$ for at least one index $\ell$.
    Nonetheless, using $|1/\Gamma(\nu_{\ell_0})| = O(1)$ for each such index, arguments parallel to Lemma~\ref{lem. asymp widehat r-N-2} yield the desired assertions.
    We leave the details to the interested reader.
\end{proof}

\subsection{Proof of Theorem~\ref{thm:real-line}} \label{sec: proof real regular}

Throughout this subsection, we assume that $p\in\R\setminus\{0\}$.
We introduce the modified prekernel
\begin{equation} \label{eq. def kappa hat}
    \widehat{\kappa}_N(z, w) := \frac{\kappa_N(z, w)}{f_1(\zeta, \eta^*) f_2(\zeta, \eta^*)}, 
\end{equation}
where $\zeta$ and $\eta^*$ are given in \eqref{eq: zeta, eta, eta-star notation}, $\kappa_N$ is defined in \eqref{eq. kappa (scaled prekernel)}, and $f_1$ and $f_2$ are defined in \eqref{eq: def f1 auxiliary func} and \eqref{eq: def f2 auxiliary func}.
Since $f_1$ and $f_2$ are holomorphic in their first argument and antiholomorphic in their second, $\widehat{\kappa}_N$ is holomorphic in both $z$ and $w$.

\begin{lemma} \label{Lem. unif bdd rescaled prekernel kappa hat}
    The sequence of prekernels $\{ \widehat{\kappa}_N(z,w)\}_{N \in \mathbb{N}}$ is locally uniformly bounded in $\C\times\C$.
\end{lemma}
\begin{proof}
    By Lemma \ref{lem. bound of varkappa}, we have
    \begin{align*}
        |\widehat{\kappa}_N(z,w)| &= \bigg| \frac{\varkappa_N(\zeta,\eta)}{(N^{1-m}\rho(p))^{3/2} f_1(\zeta,\eta^*) f_2(\zeta,\eta^*)} \bigg|
        \\
        &\leq  \frac{C}{(N^{1-m}\rho(p))^{3/2} \prod_{l=1}^m 4^{\nu_l}} 
        \bigg|\frac{U_{N,\vnu}^p(z, z) U_{N,\vnu}^p(w,w)}{f_1(\zeta,\eta^*) f_2(\zeta,\eta^*)} \bigg|.
    \end{align*}
    Here and below, $C>0$ denotes a constant independent of $N$, which may vary from line to line.
    Using \eqref{eq. asymp Meijer G} and \eqref{eq: def U-N-vnu-p}, we have
    \begin{equation*}
        U_{N,\vnu}^p(z, z) \leq C \, N^{\frac{3(1-m)}{4}} 2^{\sum_{l=1}^m\nu_l} \omega_{\vnu}(\zeta)^{-\frac{1}{2}},
    \end{equation*}
    and an analogous estimate for $U_{N,\vnu}^p(w,w)$.
    Putting them together, we have
    \begin{equation*}
        |\widehat{\kappa}_N(z,w)| \leq 
            \frac{C}{|f_1(\zeta,\eta^*) f_2(\zeta,\eta^*) \sqrt{\omega_{\vnu}(\zeta) \omega_{\vnu}(\eta)}|}.
    \end{equation*}
    Then, the assertion follows by Lemma \ref{lem. weight function and f1 f2}, noting $\eta^* = \cconj{\eta}$ for $p \in \R$ so that $\omega_{\vnu}(\eta) = \omega_{\vnu}(\eta^*)$.
\end{proof}

\begin{lemma} \label{lem: partial-kappa to partial-kappa-hat real}
    For any fixed positive integer $j$, we have as $N\to\infty$ that
    \begin{equation} \label{eq. partial-zeta-kappa to partial-z-kappa varying param}
        \zeta^j \partial_\zeta^j \kappa_N(z,w) = f_1(\zeta,\eta^*) f_2(\zeta,\eta^*) \sum_{u=0}^{j} \binom{j}{u}\left((2\Upsilon_{\vnu}(\zeta,\eta^*))^{j-u} + O(N^{j-u-1}) \right) \left(\sqrt{N\rho(p)}\,p\euler^{-\iunit\theta} +z\right)^u\partial_z^u\widehat{\kappa}_N(z,w),
    \end{equation}
    locally uniformly for $z, w \in \C$.
\end{lemma}
\begin{proof}
    Taking derivatives on \eqref{eq. def kappa hat}, direct computations show that
    \begin{align*}
        \zeta^j \partial_\zeta^j \kappa_N(z,w) &= \sum_{\substack{s,t,u \geq 0,\\ s+t+u=j}} \binom{j}{s,t,u} \big(\zeta^s\partial_{\zeta}^s f_1(\zeta,\eta^*)\big) \, \big(\zeta^t\partial_{\zeta}^t f_2(\zeta,\eta^*) \big)  \, \big(\zeta^u\partial_{\zeta}^u \widehat{\kappa}_N(z,w) \big).
    \end{align*}
    Since $\zeta = \xi_N^p(z)$, we have $\frac{\partial z}{\partial \zeta} = N^{\frac{1-m}{2}} \rho(p)^{\frac{1}{2}} \euler^{-\iunit\theta}$ which implies
    \begin{equation*}
        \zeta^k \partial_\zeta^k =  \left(\sqrt{N\rho(p)}\,p\euler^{-\iunit\theta} +z\right)^k  \partial_z^k
    \end{equation*}
    for any nonnegative integer $k$.
    Therefore, using \eqref{eq. derivative f1 var param} and \eqref{eq. derivative f2 var param}, we obtain
    \begin{align*}
        \zeta^j \partial_\zeta^j \kappa_N(z,w) &=\sum_{\substack{s,t,u \geq 0,\\ s+t+u=j}} \binom{j}{s,t,u} \left((2\Upsilon_{\vnu}(\zeta,\eta^*))^t+O(N^{t-1})\right) O(1)^s
        \\
        & \qquad\qquad \times f_1(\zeta,\eta^*) f_2(\zeta,\eta^*) \left(\sqrt{N\rho(p)}\,p\euler^{-\iunit\theta} +z\right)^u  \partial_z^u\widehat{\kappa}_N(z,w) 
        \\
        &= f_1(\zeta,\eta^*) f_2(\zeta,\eta^*) \sum_{u=0}^{j} \binom{j}{u}\left((2\Upsilon_{\vnu}(\zeta,\eta^*))^{j-u} + O(N^{j-u-1}) \right) \left(\sqrt{N\rho(p)}\,p\euler^{-\iunit\theta} +z\right)^u \partial_z^u\widehat{\kappa}_N(z,w),
    \end{align*}
    which is the desired result.
\end{proof}

\begin{lemma} \label{lem. diff eq widehat kappa (1st order apprx)}
    As $N \to \infty$, we have
    \begin{align}
        2(w-z) \widehat{\kappa}_N(z,w) +  \partial_z \widehat{\kappa}_N(z,w) + \sum_{j=0}^m \frac{a_{N,j}(z,w)}{N^{\max(1,j-1)/2}} \partial_z^j \widehat{\kappa}_N(z,w) =  \widehat{I}_{N,1}(z,w) + \widehat{I}_{N,2}(z,w) + \widehat{I}_{N,3}(z,w),
    \end{align}
    where $a_{N,j}(z,w)$ are locally uniformly bounded functions.
\end{lemma}
\begin{proof}
    Throughout the proof, we denote by $f_1 = f_1(\zeta, \eta^*)$, $f_2 = f_2(\zeta, \eta^*)$, and $\Upsilon_{\vnu} = \Upsilon_{\vnu}(\zeta, \eta^*)$.
    In addition, since the explicit expression of $a_{N,j}(z,w)$ is not of interest, we abuse notation and use $a_{N,j}(z,w)$ to denote locally uniformly bounded functions which might be different from line to line.
    
    By \eqref{eq. kappa (scaled prekernel)} and \eqref{eq. diff eq finiteN-general case}, we have
    \begin{equation} \label{eq. diff eq widehat step 1}
        \frac{1}{\zeta}\Bigg[ \prod_{l=1}^m\Big(\zeta\partial_\zeta +2\nu_l\Big) - 2^m \zeta^2\Bigg] \kappa_N(z,w) = \euler^{-\iunit\theta}\frac{ I_{N,1}(\zeta,\eta) + I_{N,2}(\zeta,\eta) + I_{N,3}(\zeta,\eta) }{(N^{1-m}\rho(p))^{3/2}}.
    \end{equation}
    Let us denote by $A$ the left-hand side of \eqref{eq. diff eq widehat step 1}. Using the expansion \eqref{eq. Meijer G diff operator expansion}, we write
    \begin{align} \label{eq: expression A}
        A= \frac{1}{\zeta}\Bigg[ \sum_{j=0}^m \syme_{m-j}(2\vnu) \zeta^{j} \partial_\zeta^j + \sum_{j=1}^{m-1} \sum_{l=0}^{m-j-1} S(m-l,j) \syme_l(2\vnu) \zeta^j \partial_\zeta^j - 2^m \zeta^2  \Bigg] \kappa_N(z,w).
    \end{align}
    For the first summation, it follows from \eqref{eq. partial-zeta-kappa to partial-z-kappa varying param} that
    \begin{align*}
        &\sum_{j=0}^m \syme_{m-j}(2\vnu) \zeta^{j} \partial_\zeta^j \kappa_N(z,w)
        \\
        &= f_1 f_2 \sum_{j=0}^m\sum_{u=0}^{j} \binom{j}{u}\syme_{m-j}(2\vnu) \left( (2\Upsilon_{\vnu})^{j-u}+O(N^{j-u-1}) \right)\left(p\sqrt{N\rho(p)} \euler^{-\iunit\theta}  +z\right)^u \partial_z^u \widehat{\kappa}_N(z,w).
    \end{align*}
    We further simplify this expression using the identities
    \begin{align*}
        \sum_{j=0}^m \syme_{m-j}(2\vnu) \, x^j
        =\syme_{m}(x+2\vnu),
        \qquad
        \sum_{j=u}^m \binom{j}{u}  \syme_{m-j}(2\vnu)\, x^{j-u} = \syme_{m-u}\left( x+2\vnu \right),
    \end{align*}
    which gives rise to
    \begin{align*}
        \sum_{j=0}^m \syme_{m-j}(2\vnu) \zeta^{j} \partial_\zeta^j \kappa_N(z,w) &= f_1 f_2 \sum_{u=0}^{m} \left(\syme_{m-u} (2\Upsilon_{\vnu}+2\vnu) + O(N^{m-u-1}) \right) \left(p\sqrt{N\rho(p)}  \euler^{-\iunit\theta}+z\right)^u \partial_z^u \widehat{\kappa}_N(z,w).
    \end{align*}
    Hence, using \eqref{eq: def Upsilon nu} and \eqref{eq: partial-Upsilon asymp}, we obtain
    \begin{align*}
    \begin{split}
        &\sum_{j=0}^m \syme_{m-j}(2\vnu)  \zeta^{j} \partial_\zeta^j \kappa_N(z,w)
        \\
        &= f_1f_2 2^m\zeta\left[\cconj{\eta^*} \widehat{\kappa}_N(z,w) + \cconj{\eta^*}\sum_{\ell=1}^{m} \frac{p\sqrt{\rho(p) N}  \euler^{-\iunit\theta}}{2\Upsilon_{\vnu}+2\nu_\ell } \partial_z \widehat{\kappa}_N(z,w)  + N^{\frac{m}{2}}\sum_{u=0}^{m}  \frac{a_{N,u}(z,w)}{N^{\max(1, u/2)}} \partial_z^u \widehat{\kappa}_N(z,w) \right],
    \end{split}
    \end{align*}
    where $a_{N,u}(z,w)$ are uniformly bounded functions defined by the equality.
    This gives rise to
    \begin{equation}
    \begin{split} \label{eq: diff eq simplification eq1}
        &\frac{1}{\zeta}\left[\sum_{j=0}^m \syme_{m-j}(2\vnu)  \zeta^{j} \partial_\zeta^j - 2^m \zeta^2\right]\kappa_N(z,w)  
        \\
        &=  2^m f_1 f_2 \Bigg[ (\cconj{\eta^*} - \zeta) \widehat{\kappa}_N(z,w) + \cconj{\eta^*} \sum_{l=1}^m \frac{p \sqrt{N\rho(p)} \euler^{-\iunit\theta}}{2\Upsilon_{\vnu}+ 2\nu_l} \partial_z \widehat{\kappa}_N (z,w) + N^{\frac{m}{2}}\sum_{u=0}^{m}  \frac{a_{N,u}(z,w)}{N^{\max(1, u/2)}} \partial_z^u \widehat{\kappa}_N(z,w)\Bigg],
    \end{split}
    \end{equation}
    where, by abuse of notation, the function $a_{N,u}(z,w)$ is locally uniformly bounded.
    
    We now turn to the analysis of the remainder which is the double summation in \eqref{eq: expression A}, and complete the proof. Again by \eqref{eq. partial-zeta-kappa to partial-z-kappa varying param}, we have
    \begin{align*}
    \begin{split} 
        &\sum_{j=1}^{m-1} \sum_{\ell=0}^{m-j-1} S(m-\ell,j) \syme_\ell(2\vnu) \zeta^j \partial_\zeta^j \kappa_N(z,w)
        \\
        &= f_1 f_2 \sum_{j=1}^{m-1} \sum_{\ell=0}^{m-j-1}\sum_{u=0}^{j} S(m-\ell,j) \syme_\ell(2\vnu) \binom{j}{u} \left( (2\Upsilon_{\vnu})^{j-u} + O(N^{j-u-1}) \right) (p\sqrt{N\rho(p)} \euler^{-\iunit\theta}+z)^u \partial_z^u \widehat{\kappa}_N(z,w).
    \end{split}
    \end{align*}
    Each summand is bounded by $O(N^{m-1-u/2})$, since $\nu_\ell$ and $\Upsilon_{\vnu}$ are of order at most $N$.
    Therefore, substituted into \eqref{eq: expression A}, they can be absorbed into $a_{N,u}(z,w)$ arising from \eqref{eq: diff eq simplification eq1}. Thus, we obtain
    \begin{equation} \label{eq: diff eq simplification eq2}
        A= 2^m f_1 f_2 \Bigg[ (\cconj{\eta^*} - \zeta) \widehat{\kappa}_N(z,w) + \cconj{\eta^*} \sum_{\ell=1}^m \frac{p \sqrt{N\rho(p)} \euler^{-\iunit\theta}}{2\Upsilon_{\vnu} + 2\nu_\ell} \partial_z \widehat{\kappa}_N (z,w) + N^{\frac{m}{2}} \sum_{u=0}^{m} \frac{a_{N,u}(z,w)}{N^{\max(1, u/2)}} \partial_z^u \widehat{\kappa}_N(z,w)\Bigg],
    \end{equation}
    where $a_{N,u}(z,w)$ denotes locally uniformly bounded functions not necessarily identical to those in \eqref{eq: diff eq simplification eq1}.
    Since $p = \cconj{p}$, it reduces to
    \begin{align*}
         A = \frac{2^{m-1} f_1 f_2}{\sqrt{N^{1-m} \rho(p)}} \euler^{-\iunit \theta} \Bigg[ 2 (w -z) \widehat{\kappa}_N(z,w) + \frac{\cconj{\eta^*}}{ N^{m/2}} \sum_{\ell=1}^m \frac{p N \rho(p) }{\Upsilon_{\vnu}+\nu_\ell} \partial_z \widehat{\kappa}_N(z,w) + \sum_{j=0}^m \frac{a_{N,j}(z,w)}{N^{\max(1, j-1)/2}} \partial_z^j \widehat{\kappa}_N(z,w) \Bigg].
    \end{align*}
    Since $\rho(p) = ( |p|^2 \sum_{\ell=1}^m \frac{N}{\Upsilon_{\vnu} + \nu_\ell})^{-1} + O(N^{-1/2})$, the conclusion follows from \eqref{eq. diff eq widehat step 1} after dividing by the prefactor.
\end{proof}

We are ready to prove Theorem~\ref{thm:real-line}.

\begin{proof}[Proof of Theorem \ref{thm:real-line}]
    We proved through Lemmas \ref{lem. asymp widehat r-N-1}, \ref{lem. asymp widehat r-N-2}, \ref{lem. asymp widehat r-N-3}, and \ref{lem. diff eq widehat kappa (1st order apprx)} that $\widehat{\kappa}_N$ satisfies a differential equation
    \begin{equation} \label{eq. proof of Thm 1 step 1}
        2  (w-z) \widehat{\kappa}_N(z,w)  +   \partial_z \widehat{\kappa}_N(z,w)  + \sum_{k=0}^m \frac{a_{N,k}(z,w)}{N^{\max(1,k-1)/2}} \partial_z^k \widehat{\kappa}_N(z,w) = \varphi(z,w) + o(1),
    \end{equation}
    as $N \to \infty$, where $a_{N,k}(z,w)$ are locally uniformly bounded functions and
    \begin{equation*}
        \varphi(z,w):= \begin{dcases}
            \frac{4}{\pi} , & \quad \text{if} \quad  |p| \in (r_{\rm in},r_{\rm out}),
            \\
            \frac{2}{\pi}\erfc\left(z+w\right) - \frac{\sqrt{2}}{\pi} \euler^{(z-w)^2-2z^2} \erfc\left(\sqrt{2}w\right), &  \quad \text{if} \quad  |p| =r_{\rm out}\quad \text{or} \quad |p|=r_{\rm in} > 0,
            \\
            0, & \quad  \text{if} \quad |p| \in (0, \infty) \setminus [r_{\rm in}, r_{\rm out}].
        \end{dcases}
    \end{equation*}

    By Lemma \ref{Lem. unif bdd rescaled prekernel kappa hat}, the sequence $\{ \widehat{\kappa}_N \}_{N \in \mathbb{N}}$ is locally uniformly bounded. Thus, by Montel's theorem, there is a convergent subsequence $\{ \widehat{\kappa}_{N_j} \}_{j \in \mathbb{N}}$, whose limit we denote by $\widehat{\kappa} := \lim_{j \to \infty} \widehat{\kappa}_{N_j}$.
    Then, \eqref{eq. proof of Thm 1 step 1} and anti-symmetry of $\widehat{\kappa}$ imply
    \begin{equation} \label{eq. proof of Thm 1 step 2}
        \begin{dcases}
            2 (w-z) \widehat{\kappa}(z,w)  + \partial_z \widehat{\kappa}(z,w)  =  \varphi(z, w),
            \\
            \widehat{\kappa}(z,z) = 0.
        \end{dcases}
    \end{equation}
    By uniqueness of the solution for initial value problems, the solution of \eqref{eq. proof of Thm 1 step 2} is uniquely given by
    \begin{equation} \label{eq: solution of kappa hat}
        \widehat{\kappa}(z,w) = \begin{dcases}
            \frac{2}{\sqrt{\pi}} \euler^{(z-w)^2} \erf(z-w), & \quad \text{if} \quad  |p| \in (r_{\rm in},r_{\rm out}),
            \\
            \frac{2}{\pi} \int_{-\infty}^0 \euler^{-s^2} \sinh(2s(w-z)) \erfc(z+w-s) \,\dif s, & \quad \text{if} \quad  |p| =r_{\rm out}\quad \text{or} \quad |p|=r_{\rm in} > 0,
            \\
            0, & \quad  \text{if} \quad |p| \in (0, \infty) \setminus [r_{\rm in}, r_{\rm out}],
        \end{dcases}
    \end{equation}
    for which we refer to \cite{BE23, Kan02, ABK22} for the detailed derivation of the solutions.
    This implies the subsequential limit is unique, and thus we have $\widehat{\kappa} = \lim_{N\to\infty}\widehat{\kappa}_N$, locally uniformly in $\C \times \C$.

    Finally, we reconstruct the correlation functions using \eqref{eq: n-point function real line}.
    Combining Lemma~\ref{lem. weight function and f1 f2}, \eqref{eq. def kappa hat} and \eqref{eq: solution of kappa hat}, we obtain the limiting kernels in Theorem~\ref{thm:real-line} for all $p \in \R \setminus \{0\}$.
\end{proof}

\subsection{Proof of Theorem~\ref{thm:complex-line}} \label{sec: proof complex regular}

Throughout this subsection, we assume that $p\in\C\setminus\R$. As in the real regime, we introduce modified prekernels
\begin{align} \label{eq. def of kappa hat ab}
    \widehat{\kappa}_N^{1,b}(z,w) := \frac{\sqrt{N\rho(p)} (\overline{p}-p) \kappa_N^{1, b}(z,w)}{f_1(\zeta,\eta_b) f_2(\zeta,\eta_b)}, 
    \qquad
    (b=1,2),
\end{align}
where $\zeta$, $\eta_1:=\eta$, and $\eta_2:=\eta^*$ are given as in \eqref{eq: zeta, eta, eta-star notation}, $\kappa_N^{1,b}$ is defined in \eqref{eq. kappa-a,b (scaled prekernel)}, and $f_1$ and $f_2$ are defined in \eqref{eq: def f1 auxiliary func} and \eqref{eq: def f2 auxiliary func}.
Notice that these functions are holomorphic in $z$.

\begin{lemma} \label{Lem. unif bdd rescaled prekernel kappa hat C}
     For each $b=1,2$, the sequence $\{ \widehat{\kappa}_N^{1,b}(z,w)\}_{N \in \mathbb{N}}$ is locally uniformly bounded in $\C\times\C$.
\end{lemma}
\begin{proof}
    It follows from \eqref{eq. bound of varkapp away from R} that
    \begin{equation*}
        |\widehat{\kappa}_N^{1,b}(z,w)| \leq \frac{C \sqrt{\rho(p)}  |\cconj{p} - p|}{(N^{1-m}\rho(p))^{3/2} \prod_{\ell=1}^m 4^{\nu_\ell}} 
        \bigg|\frac{U_{N,\vnu}^p(z, z) U_{N,\vnu}^p(w_b,w_b)}{f_1(\zeta,\eta_b) f_2(\zeta,\eta_b)} \bigg|
    \end{equation*}
    for some constant $C>0$, where $w_1:=w$ and $w_2 :=\cconj w$, which is analogous to the upper bound obtained for $\widehat{\kappa}_N$ in the proof of Lemma~\ref{Lem. unif bdd rescaled prekernel kappa hat}.
    Following the argument therein, the conclusion follows.
\end{proof}

\begin{lemma} \label{lem: partial-kappa to partial-kappa-hat complex}
    For any fixed positive integer $j$ and $b=1,2$, we have as $N\to\infty$ that
    \begin{equation*}
        \zeta^j \partial_\zeta^j \kappa_N^{1,b}(z,w) = \frac{f_1(\zeta,\eta_b) f_2(\zeta,\eta_b)}{\sqrt{N\rho(p)} (\cconj{p} - p)} \sum_{u=0}^{j} \binom{j}{u}\left( (2\Upsilon_{\vnu}(\zeta,\eta_b))^{j-u} + O(N^{j-u-1}) \right) \left(\sqrt{N\rho(p)}\,p\euler^{-\iunit\theta} +z\right)^u\partial_z^u\widehat{\kappa}_N^{1,b}(z,w),
    \end{equation*}
    locally uniformly for $z,w \in \C$.
\end{lemma}
\begin{proof}
    A similar argument to the proof of Lemma~\ref{lem: partial-kappa to partial-kappa-hat real} proves the case $p \in \C \setminus \R$ for $\kappa_N^{1,b}$ and $\widehat{\kappa}_N^{1,b}$.
\end{proof}

\begin{lemma} \label{Lem. diff eq rescaled prekernel kappa hat C}
    As $N\to\infty$, we have
    \begin{align}
        &\widehat{\kappa}_N^{1,b}(z,w)  + \sum_{k=0}^m \frac{a_{N,k}(z,w)}{N^{\max(1,k)/2}} \partial_z^k \widehat{\kappa}_N^{1,b}(z,w) = \varphi_{N}^{1,b}(z,w),
    \end{align}
    where $a_{N,k}(z,w)$ are locally uniformly bounded, and
    \begin{equation}
        \varphi_N^{1,b}(z,w) = \begin{dcases}
            \frac{1}{2}\Big(\widetilde{I}_{N,1}(z,w) + \widetilde{I}_{N,2}(z,w) + \widetilde{I}_{N,3}(z,w) \Big), & \text{if} \quad b=1,
            \\
            \frac{1}{2}\Big(\widehat{I}_{N,1}(z,w) + \widehat{I}_{N,2}(z,w) + \widehat{I}_{N,3}(z,w) \Big), & \text{if} \quad b=2.
        \end{dcases}
    \end{equation}
\end{lemma}
\begin{proof}
    Following the proof of Lemma~\ref{lem. diff eq widehat kappa (1st order apprx)}, we have
    \begin{equation*}
    \begin{split}
        &\frac{1}{\zeta}\left[\prod_{\ell=1}^m (\zeta \partial_\zeta + 2\nu_\ell) -  2^m \zeta^2\right]\kappa_N^{1,b}(z,w)  
        \\
        &=  \frac{2^m f_1(\zeta,\eta_b) f_2(\zeta,\eta_b)}{\sqrt{N \rho(p)} (\overline{p}- p)} \Bigg[ (\cconj{\eta_b} - \zeta) \widehat{\kappa}_N^{1,b}(z,w) + N^{\frac{m}{2}}\sum_{u=0}^{m} \frac{a_{N,u}(z,w)}{N^{\max(\frac{1}{2},\frac{u}{2})}} \partial_z^u \widehat{\kappa}_N^{1,b}(z,w)\Bigg],
    \end{split}
    \end{equation*}
    which is an analogue of \eqref{eq: diff eq simplification eq2}.
    Here, $a_{N,u}(z,w)$ denote locally uniformly bounded functions, which may be different from line to line.
    In contrast to Lemma~\ref{lem. diff eq widehat kappa (1st order apprx)}, since $p \neq \cconj{p}$, only the term $(\cconj{\eta_b} - \zeta) \widehat{\kappa}_N^{1,b}(z,w)$ contributes to the leading asymptotics of the expression.
    The conclusion then follows after straightforward manipulations.
\end{proof}

We are now ready to complete the proof of Theorem \ref{thm:complex-line}.

\begin{proof}[Proof of Theorem \ref{thm:complex-line}]
    We first compute the large-$N$ limit of $\widehat{\kappa}_N^{a,b}(z,w)$.
    It is enough to consider the case $a=1$, since the case $a=2$ immediately follows by complex conjugation:
    \begin{equation*}
        \kappa_N^{2,2}(\cconj{z},\cconj{w})=\cconj{\kappa_N^{1,1}(z,w)},
        \qquad
        \kappa_N^{2,1}(\cconj{z},w)=\cconj{\kappa_N^{1,2}( z,\cconj{w})}.
    \end{equation*}
    Fix any compact set $K_1 \times K_2 \subset \C\times\C$, and let $K_1' := \{z \in \C : \operatorname{dist}(z, K_1) \leq 1\}$.
    By Lemma \ref{Lem. unif bdd rescaled prekernel kappa hat C}, there exist $N_0 \in \mathbb{N}$ and $C > 0$ such that $|\widehat{\kappa}_N^{1,b}(z, w)| \leq C$ for all $(z, w) \in K_1' \times K_2$ and $N \geq N_0$, while $z \mapsto \widehat{\kappa}_N^{1,b}(z, w)$ is holomorphic on a neighborhood of $K_1'$ for each fixed $w \in K_2$.
    Applying Cauchy's integral formula on the circle $\{\widetilde{z}:|\widetilde{z} - z| = 1\}$ (with $w \in K_2$ held fixed), we obtain
    $$|\partial_z^k \widehat{\kappa}_N^{1,b}(z,w)| \leq k! C$$
    uniformly for $(z,w) \in K_1 \times K_2$ and $N \geq N_0$.
    Substituting this uniform bound into Lemma \ref{Lem. diff eq rescaled prekernel kappa hat C} together with Lemmas \ref{lem. asymp widehat r-N-1}, \ref{lem. asymp widehat r-N-2} and \ref{lem. asymp widehat r-N-3}, we obtain, locally uniformly for $z,w \in \C$, that
    \begin{align*}
        \lim_{N\to\infty} \widehat{\kappa}_N^{1,b}(z,w) = \varphi_{b}(z,w),
    \end{align*}
    where
    \begin{equation*}
        \varphi_1(z,w) = 0,
        \qquad
        \varphi_{2}(z,w) = \begin{dcases}
            \displaystyle \frac{2}{\pi}, & \qquad \text{if} \quad  |p| \in (r_{\rm in},r_{\rm out}),
            \\[1ex]
            \displaystyle \frac{1}{\pi}\erfc\left( z+w\right), &  \qquad \text{if} \quad  |p|=r_{\rm out} \ \text{or} \  |p|=r_{\rm in} > 0,
            \\
            0, & \qquad  \text{if} \quad |p| \in (0, \infty) \setminus [r_{\rm in}, r_{\rm out}].
        \end{dcases}
    \end{equation*}

    It remains to prove the determinantal formula in Theorem~\ref{thm:complex-line}. Using the above limits for $\widehat{\kappa}_N^{1,b}(z,w)$, it follows from \eqref{eq. rescaled n-point function C}, \eqref{eq. def of kappa hat ab} and Lemma~\ref{lem. weight function and f1 f2} that, as $N\to\infty$,
    \begin{align*}
        R_{N,n}(z_1, \ldots, z_n) = \Pf\begin{bmatrix}
        o(1) & \euler^{2 z_j \overline{z}_k - |z_j|^2 - |z_k|^2} \varphi_{2}(z_j,\cconj{z}_k) + o(1) \\
        - \euler^{2 z_k \overline{z}_j - |z_j|^2 - |z_k|^2} \varphi_{2}(z_k,\cconj{z}_j) + o(1) & o(1)
    \end{bmatrix}_{j,k=1}^n,
    \end{align*}
    locally uniformly for $z_1, \ldots, z_n\in\C$.
    Then, the conclusion follows from the standard identity relating the determinant and Pfaffian of a checkerboard-structured matrix:
    given an $n \times n$ square matrix $M$, we have
    \begin{equation*}
        \Pf\begin{bmatrix}
            0 & M_{jk}
            \\
            -M_{kj} & 0
        \end{bmatrix}_{j,k=1}^n =  (-1)^{- \frac{n(n-1)}{2}} \Pf \begin{bmatrix}
            0 & M
            \\
            -M^{T} & 0
        \end{bmatrix}
        = \det M.
    \end{equation*}
    This finishes the proof of Theorem~\ref{thm:complex-line}.
\end{proof}

\section{Scaling limit at the singular point}
\label{sec: Origin}
In this section, we prove Theorem~\ref{thm:origin} and Proposition~\ref{prop. origin reduction}. Unlike the regular scaling regimes considered in Section~\ref{sec:proofs}, the origin scaling involves no $N$-dependent change of variables. Moreover, the limiting prekernel retains its dependence on the induced parameters $\vnu=(\nu_1,\ldots,\nu_m)$.

Throughout this section, we assume that $\nu_\ell>-\frac12$ are fixed, unless otherwise stated.

\subsection{Proof of Theorem~\ref{thm:origin}} \label{sec: proof origin}
Recall the finite-$N$ prekernel $\varkappa_N$ defined in \eqref{eq:finite-prekernel}. As identified in \cite[Proposition~4.23]{Ips15}, it converges to
\begin{equation}
    \kappa_{\rm ori}^{\vnu}(z,w) := \lim_{N \rightarrow \infty } \varkappa_N(z,w) = 2^m \pi^{\frac{m}{2}-1} \sum_{k=0}^{\infty}\sum_{j=0}^{k} \frac{z^{2k+1}w^{2j} - w^{2k+1}z^{2j}}{\prod_{\ell=1}^{m} 4^{\nu_\ell} \Gamma(\nu_\ell +j +1) \Gamma(\nu_\ell +k +3/2)},
    \label{eq. def varkappa origin}
\end{equation}
locally uniformly for $z,w \in \C$.

\begin{proof}[Proof of Theorem~\ref{thm:origin}]
We first verify the locally uniform convergence \eqref{def. origin n-point corelation function} on $\C^n$. When $\nu_\ell > 0$ for all $\ell$, the weight $\omega_{\vnu}$ is continuous on $\C$, so \eqref{def. origin n-point corelation function} follows directly from \eqref{eq:finite-pfaffian} and \eqref{eq. def varkappa origin}.
For general $\nu_\ell > -1/2$, however, $\omega_{\vnu}(z)$ may diverge at the origin, cf.~\eqref{eq: tilde-G asymp x to 0}, so that $(\cconj{z}_j - z_j)\omega_{\vnu}(z_j)$ becomes singular near $z_j = 0$.
To absorb this singularity, we subtract the $(2j-1)$-th row and column from the $2j$-th row and column in \eqref{eq:finite-pfaffian} for each $j=1,\ldots,n$ and factor $(\cconj{z}_j - z_j)$ out of the $2j$-th row and column of the Pfaffian.
Since $\varkappa_N$ is entire and skew-symmetric, the resulting divided-difference entries converge locally uniformly by \eqref{eq. def varkappa origin} and Cauchy's estimates, while each prefactor becomes $(\cconj{z}_j - z_j)^2 \omega_{\vnu}(z_j)$, which is continuous on $\C$ for all $\nu_\ell > -1/2$. This establishes \eqref{def. origin n-point corelation function}.

To establish the integral representation \eqref{eq:origin-kernel}, it suffices to analyze
\begin{equation} \label{eq: def F-ori}
    F_{\rm ori}^{\vnu}(z,w) := 2^m\pi^{\frac{m}{2}-1} \sum_{k=0}^{\infty}\sum_{j=0}^{k} \frac{z^{2k+1}w^{2j}}{\prod_{\ell=1}^m  4^{\nu_\ell} \Gamma(\nu_\ell +j+1) \Gamma(\nu_\ell+k+3/2)},
\end{equation}
since $\kappa_{\rm ori}^{\vnu}(z,w)=F_{\rm ori}^{\vnu}(z,w)- F_{\rm ori}^{\vnu}(w,z)$.
Following the computation in Proposition~\ref{Prop. diff eq kappa}, we obtain
\begin{equation*}
    \Big[D - 2^mz^2 \Big]F_{\rm ori}^{\vnu}(z,w) = z I_{\rm ori}(zw),
\end{equation*}
where $D:= \prod_{\ell=1}^m (z \partial_z + 2\nu_\ell)$ and $I_{\rm ori}$ is given in \eqref{eq: def I-ori},
with $F_{\rm ori}^{\vnu}(0,w)=0$.
More generally, given an entire function $f$ with $f(0)=0$ and denoting by $\lambda\in\C$ an auxiliary parameter, we seek a holomorphic solution to
\begin{equation}\label{eq: euler equation lambda}
    \begin{dcases}
        (D-\lambda z^2)y(\lambda,z)=f(z),
        \\
        y(\lambda, 0) = 0.
    \end{dcases}
\end{equation}

We first consider the Euler equation $Dy_0=f$, corresponding to the case $\lambda = 0$. Its general solution is studied in \cite[Proposition~5.1]{GH25}. To formulate the result, let us write distinct elements in $\vnu=(\nu_1, \ldots, \nu_m)$ by $\{\nu_{j_1}, \ldots, \nu_{j_K}\}$ with $a_k$ denoting the multiplicity of $\nu_{j_k}$.
Then, the general solution is given by
\begin{equation} \label{eq: euler general sol}
    y_0(z) = \sum_{k=1}^K \sum_{n=0}^{a_k-1} C_{k,n} z^{-2\nu_{j_k}} (\log z)^n  + \int_{[0,1]^m} f(zs_1\cdots s_m)\prod_{j=1}^m s_j^{2\nu_j-1}\,\dif^m s,
\end{equation}
with some constants $C_{k,n} \in \C$, for $z \in \C \setminus \{0\}$.
Recall that $\nu_j > -1/2$ for all $j=1,\ldots, m$, which together with $f(0)=0$ ensures that the integral in \eqref{eq: euler general sol} defines an entire function vanishing at $0$.
Since $y_0$ is a holomorphic function near the origin with $y_0(0) = 0$, we deduce that
\begin{equation}\label{eq: y0 sol}
    y_0(z) = \int_{[0,1]^m} f(zs_1\cdots s_m)\prod_{j=1}^m s_j^{2\nu_j-1}\,\dif^m s
\end{equation}
is the unique solution to the Euler equation.

For general $\lambda$, we construct a solution of \eqref{eq: euler equation lambda} by assuming
\begin{equation}\label{eq: y lambda z series}
 y(\lambda,z)=\sum_{k=0}^{\infty}\lambda^k y_k(z)
\end{equation}
for some holomorphic functions $y_k(z)$.
Applying the differential operator, we obtain
\begin{equation*}
    (D-\lambda z^2)y =Dy_0+\sum_{k=0}^{\infty}\lambda^{k+1} \bigl(Dy_{k+1}-z^2y_k\bigr).
\end{equation*}
Thus, it is a solution of \eqref{eq: euler equation lambda} if we have for all $k\geq0$ that
\begin{equation*}
\begin{dcases}
    Dy_{k+1}=z^2y_k,
    \\
    y_k(0) = 0,
    \\
    Dy_0=f,
\end{dcases}
\end{equation*}
and the summation is well-defined.

We claim that \eqref{eq: y lambda z series} is a solution of \eqref{eq: euler equation lambda} if
\begin{equation}\label{eq: y-k sol}
 y_k(z)=\frac{z^{2k}}{2^{mk}(k!)^m}
 \int_{[0,1]^m}f(zt_1\cdots t_m)
 \prod_{j=1}^m t_j^{2\nu_j-1}(1-t_j^2)^k\,\dif^m t,
\end{equation}
for $k\geq 0$.
We proceed by induction on $k$.
For $k=0$, the claim \eqref{eq: y-k sol} directly follows from \eqref{eq: y0 sol}. 
Assuming the induction hypothesis for $k\geq0$, in parallel to \eqref{eq: y0 sol}, the solution of $Dy_{k+1}=z^2y_k$ is given by
\begin{align*}
\begin{split}
    y_{k+1}(z)
    &=\frac{z^{2k+2}}{2^{mk}(k!)^m}
    \int_{[0,1]^m}\int_{[0,1]^m}f(zs_1 \cdots s_m \sigma_1 \dots \sigma_m  )
    \prod_{j=1}^m s_j^{2\nu_j+2k+1}
    \sigma_j^{2\nu_j-1}(1-\sigma_j^2)^k
    \,\dif^m\sigma\,\dif^m s.
\end{split}
\end{align*}
Changing variables $\sigma_j \mapsto t_j$ via $t_j = s_j \sigma_j$, the integrals with respect to $s_j$ factorize, and each integral can be computed by elementary calculus.
This proves the induction hypothesis for $k+1$.
Indeed, since $f(0)=0$ yields $|y_k(z)| = O(|z|^{2k+1}/(2^{mk}(k!)^m))$ locally uniformly in $z \in \C$, the series in \eqref{eq: y lambda z series} converges locally uniformly for any $\lambda \in \C$ and can be differentiated term by term.
This proves the claim.

Moreover, the holomorphic solution of \eqref{eq: euler equation lambda} is unique. 
Indeed, suppose $h$ is holomorphic near the origin and satisfies
\begin{equation*}
    (D-\lambda z^2) h = 0, \qquad h(0)=0.
\end{equation*}
Let us write its series expansion by
\begin{equation*}
    h(z) = \sum_{k=0}^\infty a_k z^k.
\end{equation*}
Then, the coefficients satisfy
\begin{equation*}
    a_k \prod_{\ell=1}^m (k+2\nu_\ell) = \lambda a_{k-2}
\end{equation*}
for $k\geq1$ with $a_0 = a_{-1} =0$. Since $2\nu_l+1>0$, this implies $h\equiv 0$, and thus the solution of \eqref{eq: euler equation lambda} is unique.

Finally, substituting $f(z)=zI_{\rm ori}(zw)$ into \eqref{eq: y-k sol}, we obtain
\begin{equation}\label{eq: y-k I-ori}
 y_k(z,w)=\frac{z^{2k+1}}{2^{mk}(k!)^m}
 \int_{[0,1]^m}I_{\rm ori}(zw t_1\cdots t_m)
 \prod_{j=1}^m t_j^{2\nu_j}(1-t_j^2)^k\,\dif^m t.
\end{equation}
Summing $\lambda^ky_k$ and setting $\lambda=2^m$, the uniqueness of the solution implies $F_{\rm ori}^{\vnu}(z,w) = y(2^m,z)$.
This finishes the proof.
\end{proof}

\begin{remark}
To obtain an alternative representation we identify the multiple integral directly using
\cite[Eq.~(B.23)]{Ips15}, with upper parameters
$a_j=\nu_j+k+1$ and lower parameters $b_j=\nu_j$.
Since $a_j-b_j=k+1$, this gives
\begin{equation*}
 G_{m,m}^{m,0}\!\left(
 s\,\middle|\,
 \begin{matrix}
 \nu_1+k+1,\ldots,\nu_m+k+1\\
 \nu_1,\ldots,\nu_m
 \end{matrix}
 \right)=
 \frac{1}{(k!)^m}
 \int_{[0,1]^m}
 \delta\!\left(s-\prod_{j=1}^m\lambda_j\right)
 \prod_{j=1}^m\lambda_j^{\nu_j}(1-\lambda_j)^k
 \,\dif^m\lambda.
\end{equation*}
Applying this identity to \eqref{eq: y-k I-ori}, we can rewrite $y_k(z, w)$ as 
\begin{equation*}
 y_k(z,w)=\frac{z^{2k+1}}{2^{m(k+1)}}
 \int_0^1
 G_{m,m}^{m,0}\!\left(
 s\,\middle|\,
 \begin{matrix}
 \nu_1+k+1,\ldots,\nu_m+k+1\\
 \nu_1,\ldots,\nu_m
 \end{matrix}
 \right)
 I_{\rm ori}(zw\sqrt{s})\,\frac{\dif s}{\sqrt{s}}.
\end{equation*}
Summing $\lambda^ky_k$ and setting $\lambda=2^m$ now yields
\begin{align*}
    F_{\rm ori}^{\vnu}(z,w) 
    =y(\lambda=2^m,z) =\frac1{2^m}\sum_{k=0}^{\infty}z^{2k+1}
 \int_0^1
 G_{m,m}^{m,0}\!\left(
 s\,\middle|\,
 \begin{matrix}
 \nu_1+k+1,\ldots,\nu_m+k+1\\
 \nu_1,\ldots,\nu_m
 \end{matrix}
 \right)
 I_{\rm ori}(zw\sqrt s)\,\frac{\dif s}{\sqrt s}.
 \end{align*}
\end{remark}

\subsection{Reduction formula}

At the origin, the limiting prekernel $\kappa_{\rm ori}^{\vnu}$, or equivalently the $n$-point correlation function $R_n^{\vnu}$, satisfies a reduction formula analogous to that for products of complex Ginibre matrices \cite[Proposition~4.11]{Ips15}.
\begin{proposition}
    \label{prop. origin reduction}
    Let $m\geq 2$ and $n$ be a positive integer and $\hat\vnu=(\nu_1, \dots, \nu_{m-1})$ be fixed.
    Then,
    \begin{equation}\label{eq: prekernel origin reduction}
        \lim_{\nu_m \to \infty}\nu_m^{3/2}\Gamma(2\nu_m) \, \kappa_{\rm ori}^{\vnu}( \sqrt{\nu_m}z, \sqrt{\nu_m}w )
        = \kappa_{\rm ori}^{\hat\vnu}(z,w),
    \end{equation}
    locally uniformly for $z,w \in \C$. Consequently,
    \begin{equation}
        \lim_{\nu_m \to \infty }\nu_m^n R_n^{\vnu}(\sqrt{\nu_m }z_1, \ldots, \sqrt{\nu_m} z_n) = 
         R_{n}^{\hat\vnu}(z_1, \ldots,z_n),
         \label{eq: n-point origin reduction}
    \end{equation}
    locally uniformly for all $z_1, \ldots, z_n \in \Cset$.
\end{proposition}

\begin{lemma} \label{lem: weight func asymp large nu-m}
    Let $m\geq2$ and $\hat\vnu = (\nu_1, \ldots,\nu_{m-1})$ be fixed. Then, as $\nu_m \rightarrow \infty$, we have
    \begin{equation}
        \frac{\omega_{\vnu}(\sqrt{\nu_m} z)}{\Gamma(2\nu_m) \omega_{\hat{\vnu}}(z) } = 1 + O(\nu_m^{-1})
    \end{equation}
    locally uniformly for $z \in \C$.
    Here, at $z=0$, the left-hand side is understood as the continuous extension.
\end{lemma}
\begin{proof}
    Let $q\geq1$ be a fixed integer and $b_1, \ldots, b_q > -1$ be fixed real numbers. We claim for any $M>0$ that there is a constant $C_M>0$ such that, as $b\to\infty$,
    \begin{equation} \label{eq: weight func asymp large nu-m claim}
        \Bigg|\frac{1}{\Gamma(b) \, \widetilde{G}(x)} G_{0,q+1}^{q+1,0}\Big( b x \Big| \genfrac{}{}{0pt}{}{-}{b_1, \ldots, b_q, b} \Big) -1 \Bigg| \leq \frac{C_M}{b}
    \end{equation}
    for all $0<x\leq M$, where
    \begin{equation}
        \widetilde{G}(x):= G_{0,q}^{q,0}\Big( x \Big| \genfrac{}{}{0pt}{}{-}{b_1, \ldots, b_q} \Big).
    \end{equation}
    Once the claim is proved, the conclusion of the lemma directly follows upon substituting $q= m-1$, $b_j=2\nu_j$ ($j=1, \ldots, m-1$), $b = 2\nu_m$, and $x = 2^{m-1}|z|^2$.

    For the rest of the proof, we denote by $C \in \R \setminus \{0\}$ and $c \in \C \setminus \{0\}$ constants, which may differ from line to line. They may depend on the parameters $q$, $b_1, \ldots, b_q$, and $M$.
    
    To estimate the derivatives of $\widetilde{G}$, let $b_{\rm min} := \min\{b_1, \ldots, b_q\}$, and let $d$ be its multiplicity.
    It is known as $x\to0$ that 
    \begin{equation} \label{eq: tilde-G asymp x to 0}
        \widetilde{G}(x) = c \, x^{b_{\rm min}} (\log x)^{d-1} \, (1+ o(1)),
    \end{equation}
    see \cite[Section 3]{GH25}. Denoting by $D:= x \partial_x$, we recall the general identity for the Meijer $G$-function \cite[Section 5.4]{Luke69}:
    \begin{equation*}
        D G_{p,q}^{m,n}\Big(x \Big|  \genfrac{}{}{0pt}{}{a_1, \ldots, a_p}{b_1, \ldots, b_q} 
        \Big) =  b_1 G_{p,q}^{m,n}\Big(x \Big|  \genfrac{}{}{0pt}{}{a_1, \ldots, a_p}{b_1, \ldots, b_q} 
        \Big) - G_{p,q}^{m,n}\Big(x \Big|  \genfrac{}{}{0pt}{}{a_1, \ldots, a_p}{b_1+1, b_2, \ldots, b_q} 
        \Big).
    \end{equation*}
    Thus, this gives rise to a rough bound
    \begin{equation*}
        D^j \widetilde{G}(x) = O(x^{b_{\rm min}} (1 + (\log x)^{d-1} )),
    \end{equation*}
    as $x \to 0$, for any fixed integer $j \geq 1$.
    Hence, we have
    \begin{equation} \label{eq: estimate (D + D2)-widetilde-G}
        \frac{|D \widetilde{G}(x)| + |D^2 \widetilde{G}(x)|}{|\widetilde{G}(x)|} \leq C
    \end{equation}
    for all $0<x \leq M$.
    On the other hand, using the Mellin convolution identity repeatedly \cite[Eq.~(B.24)]{Ips15}, we have
    \begin{equation*}
        \widetilde{G}(x) = \int_0^\infty \cdots \int_0^\infty \dif t_1 \cdots \dif t_{q-1}  \, \bigg(\prod_{\ell=1}^{q-1} t_\ell^{b_\ell-1} \euler^{-t_\ell} \bigg) \Big( \frac{x}{t_1 \cdots t_{q-1}} \Big)^{b_q} \exp\Big({- \frac{x}{t_1 \cdots t_{q-1}}}\Big) .
    \end{equation*}
    Since the integrand is positive, we deduce that
    \begin{align}
    \begin{split} \label{eq: estimate D-widetilde-G}
        D \widetilde{G}(x) &= b_q \widetilde{G}(x) -  \int_0^\infty \cdots \int_0^\infty \dif t_1 \cdots \dif t_{q-1}  \, \bigg(\prod_{\ell=1}^{q-1} t_\ell^{b_\ell-1} \euler^{-t_\ell} \bigg) \Big( \frac{x}{t_1 \cdots t_{q-1}} \Big)^{b_q} \frac{x}{t_1 \cdots t_{q-1}} \exp\Big({- \frac{x}{t_1 \cdots t_{q-1}}}\Big)
        \\
        &\leq b_q \widetilde{G}(x).
    \end{split}
    \end{align}

    We now turn to prove the claim \eqref{eq: weight func asymp large nu-m claim}.
    Using the Mellin convolution identity for the Meijer $G$-functions, we have
    \begin{equation*}
        \frac{1}{\Gamma(b) \, \widetilde{G}(x)} G_{0,q+1}^{q+1,0}\Big( b x \Big| \genfrac{}{}{0pt}{}{-}{b_1, \ldots, b_q, b} \Big) = \int_0^\infty f_b(s) g_x(s)  \,\dif s,
    \end{equation*}
    where
    \begin{equation*}
        f_b(s) := \frac{b^b}{\Gamma(b)} s^{b-1} \euler^{-bs},
        \qquad
        g_x(s) := \frac{\widetilde{G}(x/s)}{\widetilde{G}(x)}.
    \end{equation*}
    It follows from direct computations that
    \begin{equation} \label{eq: f-b integral Gamma func}
        \int_0^\infty s^a f_b(s) \,\dif s = \frac{\Gamma(a+b)}{b^a \Gamma(b)}
    \end{equation}
    for $a \geq 0$, and
    \begin{equation*}
        g_x(1) = 1,
        \qquad
        g_x'(1) = - \frac{D\widetilde{G}(x)}{\widetilde{G}(x)},
        \qquad
        g_x''(s) = \frac{1}{s^2} \frac{(D^2+D) \widetilde{G}(\frac{x}{s})}{\widetilde{G}(x)}.
    \end{equation*}
    In particular, from \eqref{eq: f-b integral Gamma func} we have
    \begin{equation} \label{eq: weight func uniform estimation int rep}
        \frac{1}{\Gamma(b) \, \widetilde{G}(x)} G_{0,q+1}^{q+1,0}\Big( b x \Big| \genfrac{}{}{0pt}{}{-}{b_1, \ldots, b_q, b} \Big) - 1 = \int_0^\infty f_b(s) \Big( g_x(s) - 1 - g_x'(1) (s-1) \Big) \,\dif s.
    \end{equation}
    
    Let $\epsilon>0$ be a sufficiently small constant. 
    It follows from \eqref{eq: estimate (D + D2)-widetilde-G} that
    \begin{equation*}
        |g_x(s)- 1 - g_x'(1) (s-1)| \leq C (s-1)^2
    \end{equation*}
    for all $|s-1|\leq \epsilon$. This shows that
    \begin{equation} \label{eq: weight func uniform estimation int rep near s=1}
        \int_{1-\epsilon}^{1+\epsilon} f_b(s) \Big|g_x(s) - 1 - g_x'(1)(s-1)\Big| \,\dif s \leq \int_0^\infty f_b(s) \, C(s-1)^2 \,\dif s \leq \frac{C}{b},
    \end{equation}
    where we used \eqref{eq: f-b integral Gamma func} for the last inequality.

    It remains to estimate the integral for $s$ bounded away from $1$.
    For $s \geq 1$, we notice that $x/s$ and $x$ are all contained in the interval $(0,M]$, from which it follows that $g_x(s) \leq s^C$ by \eqref{eq: estimate (D + D2)-widetilde-G}.
    On the other hand, for $s<1$, we have $g_x(s) \leq s^{- b_q}$ by \eqref{eq: estimate D-widetilde-G}. Combining these two with $|g_x'(1)| \leq C$ from \eqref{eq: estimate (D + D2)-widetilde-G}, we deduce that there exist constants $A\geq 1$ and $C_1>0$ such that
    \begin{equation*}
        \Big| g_x(s) - 1 - g_x'(1)(s-1) \Big| \leq C_1 (s^A + s^{-A})
    \end{equation*}
    for all $s>0$ and $x \in (0,M]$.
    Moreover, since $s\mapsto s-1 -\log s$ is a convex function attaining its minimum at $s=1$, we have
    \begin{equation*}
        f_b(s) = \frac{b^b \euler^{-b}}{\Gamma(b)} s^{-1} \euler^{-b (s-1 - \log s)} \leq \frac{b^b \euler^{-b}}{\Gamma(b)} s^{-1} \euler^{- \frac{b}{2} c_\epsilon} \euler^{-\frac{b}{2} (s-1 - \log s)},
    \end{equation*}
    for some constant $c_\epsilon>0$ and all $s \in (0, 1-\epsilon) \cup (1+\epsilon,\infty)$.
    Therefore, combining these estimates, we deduce
    \begin{align}
    \begin{split}
        &\bigg( \int_0^{1-\epsilon} + \int_{1+\epsilon}^\infty \bigg) f_b(s) \Big| g_x(s) - 1 - g_x'(1) (s-1) \Big| \, \dif s
        \\
        &\leq C_1 \frac{b^b \euler^{-b}}{\Gamma(b)} \euler^{- \frac{b}{2} c_\epsilon} \bigg( \int_0^{1-\epsilon} + \int_{1+\epsilon}^\infty \bigg)  s^{-1} \euler^{-\frac{b}{2} (s-1 - \log s)} (s^A + s^{-A}) \, \dif s
        \\
        &\leq C_1 \frac{b^b \euler^{-b}}{\Gamma(b)} \euler^{- \frac{b}{2} c_\epsilon} \int_0^\infty s^{-1} \euler^{-\frac{b}{2} (s-1 - \log s)} (s^A + s^{-A}) \, \dif s
        = O(\euler^{-b \, C'}), \label{eq: weight func uniform estimation int rep away s=1}
    \end{split}
    \end{align}
    as $b\to\infty$, for some constant $C'>0$.
    Putting \eqref{eq: weight func uniform estimation int rep}, \eqref{eq: weight func uniform estimation int rep near s=1}, and \eqref{eq: weight func uniform estimation int rep away s=1} together, the claim \eqref{eq: weight func asymp large nu-m claim} follows.
\end{proof}

\begin{proof}[Proof of Proposition~\ref{prop. origin reduction}]
    By the definition \eqref{eq. def varkappa origin}, we have
    \begin{align*}
        \nu_m^{3/2}\Gamma(2\nu_m) \kappa_{\rm ori}^{\vnu} (\sqrt{\nu_m}\,z,\sqrt{\nu_m}\,w) 
        = 2^m\pi^{\frac m2-1} \sum_{k=0}^\infty\sum_{j=0}^k \frac{z^{2k+1}w^{2j}-w^{2k+1}z^{2j}}{\prod_{\ell=1}^{m-1} 4^{\nu_\ell} \Gamma(\nu_\ell+j+1) \Gamma(\nu_\ell+k+\frac32)} \,A_{j,k}(\nu_m),
    \end{align*}
    where
    \begin{equation*}
        A_{j,k}(\nu) := \frac{ \nu^{j+k+2}\Gamma(2\nu) }{ 4^\nu \Gamma(\nu+j+1) \Gamma(\nu+k+\frac{3}{2}) }.
    \end{equation*}
    
    For every fixed $j,k\geq0$, Stirling's formula and the duplication formula give rise to
    \begin{equation*}
        \lim_{\nu \to \infty} A_{j,k}(\nu) = \frac{1}{2\sqrt{\pi}}.
    \end{equation*}
    Moreover, for any $x\geq1$ and $j\geq0$, we have $\Gamma(x+j+1) \geq x^j \Gamma(x+1)$, which implies
    \begin{equation*}
        0<A_{j,k}(\nu) \leq \frac{\nu^2 \Gamma(2\nu)}{4^\nu \Gamma(\nu+1) \Gamma(\nu+\frac{3}{2})} \leq C,
    \end{equation*}
    for some constant $C>0$, uniformly for $j,k \geq 0$ and sufficiently large $\nu$.
    Since $\nu_1, \ldots, \nu_{m-1}$ are fixed, the double summation
    \begin{equation*}
        \sum_{k=0}^\infty \sum_{j=0}^k \frac{z^{2k+1}w^{2j}}{\prod_{\ell=1}^{m-1}4^{\nu_\ell}\Gamma(\nu_\ell+j+1) \Gamma(\nu_\ell+k+\frac{3}{2})}
    \end{equation*}
    is absolutely convergent. Hence, using the bound for $A_{j,k}(\nu_m)$, the Weierstrass $M$-test implies
    \begin{align*}
        \lim_{\nu_m \to \infty} \nu_m^{3/2}\Gamma(2\nu_m) \kappa_{\rm ori}^{\vnu} (\sqrt{\nu_m}\,z,\sqrt{\nu_m}\,w)
        &= 2^{m-1} \pi^{\frac{m-3}{2}} \sum_{k=0}^\infty \sum_{j=0}^k \frac{z^{2k+1}w^{2j} - w^{2k+1}z^{2j}}{\prod_{\ell=1}^{m-1}4^{\nu_\ell}\Gamma(\nu_\ell+j+1) \Gamma(\nu_\ell+k+\frac{3}{2})}
        = \kappa_{\rm ori}^{\hat{\vnu}}(z,w),
    \end{align*}
    which proves \eqref{eq: prekernel origin reduction}.
    Together with Lemma~\ref{lem: weight func asymp large nu-m} and \eqref{def. origin n-point corelation function}, using the same row and column subtraction as in the proof of Theorem~\ref{thm:origin}, this proves \eqref{eq: n-point origin reduction}.
    This finishes the proof.
\end{proof}

\end{document}